\documentstyle[preprint,epsfig,aps]{revtex} 
\tightenlines
\def\lsim{\mathrel{\rlap{\lower4pt\hbox{\hskip1pt$\sim$}}
    \raise1pt\hbox{$<$}}}         
\def\gsim{\mathrel{\rlap{\lower4pt\hbox{\hskip1pt$\sim$}}
    \raise1pt\hbox{$>$}}}         
\def\freg{f_{\mbox{\rm reg}}}    
\begin{document}
\begin{titlepage}
\preprint{
\vbox{ \hbox{IFIC/00-73}
         \hbox{hep-ph/0012313}
         \hbox{}\hbox{} }}
\title{
Zenith angle distributions at Super-Kamiokande and SNO and the solution
of the solar neutrino problem}

\vskip 2cm

\author{M. C. Gonzalez-Garcia \thanks{concha@flamenco.ific.uv.es}
and Carlos Pe\~na-Garay \thanks{penya@flamenco.ific.uv.es}}
\address{Instituto de F\'{\i}sica Corpuscular \\
Universitat de  Val\`encia -- C.S.I.C\\
Edificio Institutos de Paterna, Apt 2085, 46071 Val\`encia, Spain}
\vskip 1cm
\author{Alexei Yu. Smirnov\thanks{smirnov@ictp.trieste.it}}
\address{
International Center for Theoretical Physics, 34100 Trieste, Italy\\
Institute for Nuclear Research, Russian Academy of Sciences, 
Moscow 117312, Russia} 
\maketitle

\vskip 0.5cm

\begin{abstract}

We have performed a detailed study of the zenith angle 
dependence  of the regeneration factor and 
distributions of
events at SNO and SK for different solutions of the solar neutrino 
problem. In particular, we discuss oscillatory behaviour and 
the synchronization  effect in the distribution for the LMA
solution, the parametric peak for the LOW solution, etc\dots
Physical interpretation  of the effects is given.  
We suggest a new binning of events  which emphasizes 
distinctive features of zenith angle distributions for the different
solutions.  We also find the correlations between the integrated day--night 
asymmetry and the rates of events in different zenith angle bins. 
Study  of these  correlations strengthens the 
identification power of the analysis.  
\end{abstract}

\end{titlepage}

\section{Introduction}
\label{sec:introduction}

Oscillations of solar neutrinos in the matter of the Earth modify 
the neutrino  signal detected during the 
night \cite{MS86,DN,other}. 
The integral characteristic of this  effect is the 
the day--night asymmetry: 
\begin{equation}
A_{\rm{DN}} \equiv  2\,\frac{N - D}{N + D}~, 
\label{adn}
\end{equation}
where $N$ and  $D$ are the  night and the  day event rates, 
averaged  over the year, and corrected for the Earth orbit eccentricity.  

In most of the cases, the Earth--matter
effect leads to the regeneration of the $\nu_e$ - flux, so the asymmetry
is positive. Conversely, a negative asymmetry appears for the SMA solution 
at small mixing angles. There is a number of detailed studies of the
asymmetry and its dependence on neutrino energy and oscillation
parameters \cite{MS86}-\cite{gpns}. Asymmetries have also been studied for
certain intervals of the zenith angles \cite{MPSNO}.

The observation of a day--night asymmetry will 
be the proof of the matter conversion solution of the solar neutrino problem, 
excluding the vacuum oscillation solutions. 
However, the measurement of the asymmetry alone may not allow 
to select among the three possible MSW solutions: LMA, SMA and LOW. 
A negative asymmetry  is the evidence of the SMA solution. 
However the expected value of the effect is small $|A_{\rm{DN}}| < 0.01$ and
it will be very difficult to establish experimentally such a small 
deviation from zero. Furthermore, one can  make certain conclusions 
confronting the value of the asymmetry with other solar 
neutrino data (rates, energy 
spectrum of the recoil electrons). In particular, a relatively large
asymmetry will be inconsistent with other solar neutrino 
data (rates, recoil electron energy spectrum)  in the case of SMA and LOW 
solutions, thus favoring the LMA solution. 

Further insight can be obtained by studying the zenith angle distribution 
of events during the night. It turns out that 
different solutions lead to qualitatively different distributions. 
One of the first detailed studies of the zenith angle dependence of events 
for LMA, SMA and LOW solutions of the solar neutrino problem  
was done by Baltz and Wenesser \cite{BW} and then further elaborated 
in \cite{BK,LH,LM,Rosen,lma,gpns}.

It was  realized that for the SMA solution the distribution has a
peculiar form with rather small effect for neutrinos crossing the
mantle only  ($\cos \theta_Z < 0.837$) and large 
regeneration effect (peak) for neutrinos whose
trajectories crossed both the mantle and the core of the Earth
\cite{BW,LM,Rosen,losecco}.  
This peak at large $\cos \theta_Z$ was interpreted as due to certain
constructive interference 
of the oscillation effects in the mantle and in the core of the Earth
\cite{pet} or in more simple and transparent way, as the effect of 
parametric enhancement of oscillations \cite{akh} 
(see also \cite{para_old,KS} and for later discussion \cite{CP,AkS}). 

For the LMA solution one expects averaging of oscillations 
due to integration over the energies of neutrino and 
detected charge lepton  as well as over  finite size of
the zenith angle bins. This leads to a  rather 
flat distribution with small variations of the average rate during the
night \cite{BW,lma,HPGV,GRS}. A  significant regeneration 
effect is expected already in the first 
night bin $\cos \theta_Z < 0.2$.  

For the LOW solution one gets 
the highest rate in the second night bin \cite{BW,gpns}, 
followed by a dip and then again 
an increase of the rate for large $\cos \theta_Z$. 
This peak has a simple interpretation as the oscillation maximum which
corresponds to the phase of oscillations $\phi = \pi$. 

Thus, a detailed study of the zenith angle distribution will allow to
disentangle the solutions. Moreover, it may  allow to determine the 
oscillation parameters. Measurements of the zenith angle distribution 
are also interesting because they open the possibility  to study various
matter oscillation  effects, such as: 
\begin{itemize}
\item
oscillations in matter with constant and slowly changing density, 
\item
adiabatic conversion, 
\item
effects of the resonance enhancement of oscillations, 
\item
parametric effects, 
\item
oscillation effects in thin layers (with small density width),  
\item
 effects of small density jumps, etc..
\end{itemize}

Furthermore, in principle, (if the oscillation parameters are
determined from some other experiments)  
the precise measurement of the zenith angle distribution will allow to 
check models of the Earth density profile. 

In this paper we continue to study in detail the zenith angle 
distributions for the different MSW solutions of the solar neutrino problem. 
We give physical interpretation of various features of the distributions. 
We clarify what can be learned besides identification of the solution of
the solar neutrino problem by measuring  the zenith angle
distributions in the present and in future high statistics experiments.

The paper is organized as follows: in Sec. \ref{sec:physics} we present 
 general expressions for the survival probability and the
regeneration  factor which describe the Earth--matter effect and
we study  the properties of the regeneration factor in the constant 
and slowly changing density
approximation.  In Sec.~\ref{sec:regrates} we present the results of our 
numerical calculations of the regeneration factor and the rates for the 
Super--Kamiokande and SNO for a realistic Earth--matter profile. We give an 
interpretation of the results of these calculations using the results of 
the  analytical studies. 
In Sec.~\ref{sec:bincor} we propose a new $\cos \theta_Z$ binning  which 
emphasizes  distinctive features of the distributions for the
different solutions. 
We also study the correlations of rates as well as ratios of rates with 
the value of the day--night asymmetry as a way to 
strengthen the identification power of analysis. 
We summarize and discuss our results in Sec~\ref{sec:conclusion}. 

\section{General relations}
\label{sec:physics} 

\subsection{Regeneration factor and conversion inside the Sun}
\label{subsec:regen}

For the range of oscillation parameters of interest 
($\Delta m^2 > 10^{-8}$ eV$^{2}$), the probability
$P_{ee}$ to detect 
the solar electron 
neutrino at a detector 
can be written as follows:
\begin{equation}
P_{ee}\ =\ P_1 + (1-2P_1)\left(\sin^2 \theta + \freg\right)~.   
\label{Pnig}
\end{equation}
In Eq.(\ref{Pnig}) the Earth regeneration factor, $\freg$, is defined 
as \cite{gpns}: 
\begin{equation}
\freg \equiv P_{2e}  - \sin^2 \theta~,  
\label{regen}
\end{equation}
where  $P_{2e}$ is the probability of the 
$\nu_{2} \rightarrow \nu_e$ conversion inside the Earth. 
In the absence of the Earth--matter effect we have 
$P_{2e} = \sin^2 \theta$, so that $\freg = 0$. 

In Eq.~(\ref{Pnig})  $P_1$ is the probability of the 
$\nu_{e} \rightarrow \nu_1$ conversion inside the Sun which can be  
approximated by the well known expression
\begin{equation}
P_1 = \frac{1}{2} + \left(\frac{1}{2} - P_c \right)\cos2\theta_S~.   
\label{Ponead}
\end{equation}
Here $\theta_S$ is the matter mixing angle at the production point  
inside the Sun:  
\begin{equation}
\cos 2\theta_S \equiv \cos2\theta_m(\rho_S) 
\label{eq}
\end{equation}
where $\theta_m(\rho)$ is the mixing angle in matter determined by 
\begin{equation}
\cos2\theta_m  = {-1 + \eta \cos2\theta\over(1-2\eta
\cos 2\theta+\eta^2)^{1/2}}.
\label{tSnusun}
\end{equation}
Here 
\begin{equation}
\eta \equiv \frac{l_0}{l_\nu} = 
\frac{\sqrt 2 m_N}{G_F \rho Y_e} \frac{\Delta m^2}{E} = 
0.66\left({\Delta m^2/E\over10^{-13}\ \mbox{\rm eV}}\right)
\left({ 1\ \mbox{\rm g\ cm}^{-3}\over\rho Y_e}\right) 
\label{defeta}
\end{equation}
is the ratio between the refraction length, $l_0$, and the neutrino 
oscillation length in vacuum, $l_\nu$:
\begin{eqnarray}
l_0\equiv\frac{2\pi m_N}{\sqrt2 G_F\rho Y_e},  &
& \ \ \ \ l_\nu\equiv\frac{4\pi E}{\Delta m^2}~. 
\label{defl0}
\end{eqnarray}
In Eqs.~(\ref{defeta}) and~(\ref{defl0}) 
$\rho$ is the matter density, $Y_e$
is the number of electrons 
per nucleon and $m_N$ is the nucleon mass. $P_c$ is the jump probability
which takes for an exponential density 
profile  the following form \cite{Petc,KrPe}: 
\begin{equation}
P_c = {e^{-\gamma\sin^2\theta}-e^{-\gamma}\over1-e^{-\gamma}}, 
\label{finPc}
\end{equation}
where $\gamma$ is the ratio of the density scale height $l_\rho$ and 
the neutrino oscillation length:
\begin{eqnarray}
\gamma\equiv\frac{4\pi^2 l_\rho}{l_\nu}=
1.05\left({\Delta m^2/E\over10^{-15}\ \mbox{\rm eV}}\right)
\left({l_\rho\over r_0}\right),&  & ~~~~
l_\rho\equiv{\rho\over d\rho/dr}.
\label{defgam}
\end{eqnarray}
$r_0=R_\odot/10.54$ is the height scale in the exponential
approximation to the solar density profile. 
Inserting (\ref{Ponead}) into (\ref{Pnig}) we get  
\begin{equation}
P_{ee} = P_D -  (1 - 2P_c) \cdot \cos2\theta_S \cdot \freg~,    
\label{probtot}
\end{equation}
where
\begin{equation}
P_D  = \frac{1}{2} + \frac{1}{2}(1 - 2P_c) \cdot \cos2\theta_S 
\cdot \cos2\theta 
\label{Pday}  
\end{equation}
is the survival probability in the absence of the Earth--matter effect, 
i.e.,  during the day. 

According to Eq.~(\ref{probtot}), the regeneration effect appears 
multiplied by two factors: 

1) The {\it adiabaticity factor}, $(1 - 2P_c)$, which describes the
adiabaticity
of the conversion inside the Sun. The factor is maximal for 
$P_c = 0$, that is, for the case of pure adiabatic propagation and 
it decreases with the increase of the adiabaticity breaking. 
For $P_c = 1/2$ the factor and the regeneration effect are zero,  
and at $P_c < 1/2$ the Earth--matter effect changes the sign. 

2) The {\it resonance position factor}, $\cos2\theta_S$, which determines
how far 
the resonance layer is situated from the production point (in the density 
scale).  When neutrinos are produced at the resonance,  
one has $\cos2\theta_S = 0$,  
while $\cos2\theta_S < 0$ if the resonance occurs at densities lower 
than the one at the production point. The parameter 
$\cos2\theta_S$ decreases with 
the resonance density, and it reaches the value $\cos2\theta_S \approx -1$ 
when the resonance layer is 
sufficiently far out from the production region 
(i.e.,  at much lower densities). 

From the previous discussion we conclude that the largest regeneration 
effect occurs when the neutrino propagation is adiabatic 
and the resonance happens far from the production region. 

From Eq.~(\ref{probtot}) one can obtain the daily average survival 
probability which for $\eta\ll 1$ ($\cos2\theta_S \approx -1$) 
takes the form:
\begin{equation}
\bar P\equiv{1\over2}(P_D + P_N) 
= {1\over2}[1 + (1-2P_1)(\freg-\cos2\theta)].
\label{barP}
\end{equation}

The numbers of events during the day and during the night are  
proportional to averaged probabilities $P_D$ and  $P_N$, therefore the 
day--night asymmetry (\ref{adn}) can be written as 
\begin{equation}
A_{\rm{DN}}\equiv{P_N - P_D\over\bar P}
={2\freg\over1/(1-2P_1)-\cos2\theta+\freg}~, 
\label{defAND}
\end{equation}
where both in Eq.(\ref{barP}) and Eq.(\ref{defAND}), $\freg$ should be 
averaged over the  neutrino trajectories during the night.

\subsection{Regeneration factor in constant density and adiabatic approximations}
\label{subsec:physics}

To obtain the zenith angle distributions  we have performed exact
numerical calculations 
of the regeneration factor integrating  the evolution
equation in the Earth--matter with  the Earth density profile given in the
Preliminary Reference Earth Model (PREM) \cite{PREM}. However, 
a number of qualitative features of our results can be easily 
understood in the simplified two--layers approximation of the Earth profile.
In this approximation the profile consists of the mantle and the core
with slowly changing  densities and a sharp change of the density between
the layers. In what follows we will parametrize the zenith angle 
dependence in terms of $|\cos \theta_Z|$ which, for simplicity, 
we will write omitting the moduli. 
Equivalently, this would correspond to $\theta_Z$ being the 
nadir angle.  We call the region of zenith angles 
$\cos \theta_Z = 0 - 0.837$,  
for which neutrinos cross the mantle only,  
the {\it mantle region}  and the region $\cos \theta_Z = 0.837 - 1$, 
for which neutrinos cross both the mantle and the core, 
the {\it core region}. 

Simple analytical results can be obtained in the constant density 
approximation, or, in general, in the adiabatic approximation.  
In the  case  of constant density  (which would correspond to 
neutrinos crossing the mantle  at small enough $\cos \theta_Z$ ) we obtain
the following expression for the regeneration factor 
\begin{equation}
\freg =  \sin2\theta_m \sin(2\theta_m - 2 \theta) \cdot
\sin^2\left(\frac{\pi d}{l_m}\right)\, , 
\label{regen1}      
\end{equation}
where $\theta_m$ is the mixing angle in the Earth matter  given 
in Eq.~(\ref{tSnusun}) with $\eta$ evaluated for the Earth density, 
and $d$ is the width of the layer or equivalently the distance travelled
by the neutrino. Using the expression for the mixing angle 
(\ref{tSnusun}) we get that:
\begin{equation}
\freg = 
{\eta \sin^22\theta \over (1 - 2\eta \cos2\theta + \eta^2)} 
\cdot \sin^2\left(\frac{\pi d}{l_m}\right)
\label{exp}
\end{equation}
which can be rewritten in the following form (see also \cite{BW}):
\begin{equation}
\freg = 
\frac{1}{\eta} \cdot \sin^2 2\theta_m  \cdot
\sin^2 \left(\frac{\pi d}{l_m}\right)~. 
\label{regen2} 
\end{equation}
The oscillation length in matter, $l_m$, equals 
\begin{equation}
l_m  
= l_{\nu}\frac{\sin 2\theta_m}{\sin 2\theta} 
=  l_{\nu}\frac{\eta}{(1- 2\eta \cos2\theta + \eta^2)^{1/2}}~ 
=  \frac{l_0}{(1- 2\eta \cos2\theta + \eta^2)^{1/2}}~,   
\label{l-m}
\end{equation}
where $l_0$ and $l_\nu$ are defined in Eq.(\ref{defl0}) and  
$\eta$ should be taken for the Earth density. 

Written in the form of Eq.(\ref{regen2}), $\freg$ differs from the 
expression for the probability of usual flavor oscillations in matter 
by the factor ${1}/{\eta} \propto n_e E/ \Delta m^2$. One can notice
as well that the amplitude of oscillations of $\freg$ 
\begin{equation}
A_f = \frac{1}{\eta} \sin^2 2\theta_m 
\label{depthA}
\end{equation}
is symmetric with respect to the exchange 
${\eta} \leftrightarrow {1}/ {\eta}~$, as can be seen explicitly from 
the Eq. (\ref{exp}). 
Due to the additional factor ${1}/{\eta}$ the asymptotics of the 
oscillation amplitude are  $A_f \rightarrow 0$  both at ${\eta}
\rightarrow 0$ 
and  at ${\eta} \rightarrow \infty$ which 
differ from the ones for flavor oscillations \cite{MS86}.   

According to Eq.~(\ref{regen2}), in the case of one  layer with 
constant density, the regeneration $\freg$ is always (for
any value $\Delta m^2/E$ and $\theta$)  positive, 
{\it i.e.} the matter effect of 
Earth always {\it enhances} the survival probability $P_{ee}$ 
(see Appendix).


In fact, in the mantle (and also in the core) the density 
changes rather significantly. According to PREM model  it increases 
from $\rho\sim3.2$ g/cm$^3$  near the surface  of the Earth to 
$\rho\sim 5.6$ g/cm$^3$ 
at the border with the core. The density changes smoothly apart from 
several jumps at distances $(0.05 - 0.12)R_{E}$ from the surface. 
Such a density variation leads to deviation from the simple oscillation
picture describe above. One expects certain interplay of the oscillations 
and the adiabatic evolution which results in the change 
of the oscillation probability from the constant density description. 
Furthermore the small jumps in the density 
may induce some irregularities in the
behavior of the zenith angle distribution. 

The description of the effects is simple if the adiabaticity 
condition is fulfilled 
(as it happens for the LMA solution). In this case the 
average probability and the amplitude  of oscillations are determined
uniquely by the instantaneous value of the density. Therefore 
$\bar{P}$ and the depth of oscillations of $\freg$ 
will be determined by the density at the surface $\rho_{s}$:  
\begin{equation}
\freg =
\frac{1}{\eta (\rho_{s})} \cdot \sin^2 2\theta_m(\rho_{s}) \cdot
\sin^2 \left(\pi \int \frac{dx}{l_m}\right)~.
\label{regen3}
\end{equation}

In some cases when adiabaticity is violated one can describe  the results
using constant density approximation with some effective value of the
density for each trajectory:  
\begin{equation}   
l_m = l_m(\theta_Z), ~~~ \theta_m = \theta_m(\theta_Z), ~~~ 
\eta = \eta(\theta_Z). 
\label{dep}
\end{equation}

If the oscillations are averaged due to integration 
over the energy or/and the zenith angle we get 
\begin{equation}
\bar{f}_{\rm reg} = {\eta \sin^22\theta \over 2(1 - 2\eta \cos2\theta +
\eta^2)}  
\label{fregear2}
\end{equation}
which coincides with expression used in \cite{gpns} with $\eta$ taking
an average value for the corresponding $E$ and $\theta_Z$ range. 

In the case of the core crossing trajectories the description becomes more
complicated \cite{pet,akh,CP,AkS,param_above}.

The analytical results presented in this section allow us to get a  
straightforward interpretation of the results of the numerical 
computations. Moreover,  they give the correct functional dependence 
of  observables on the neutrino parameters.

\section{Zenith angle dependence of the regeneration factors and rates}
\label{sec:regrates}

In this section we consider the zenith angle dependence
of the  survival probability in Eq.~(\ref{Pnig}),  
the regeneration factor in Eq.~(\ref{regen}),
and the event rates at SK and SNO for oscillation 
parameters from  the LMA, LOW, and SMA solutions to the solar neutrino 
problem. 

The reduced rate of the electron scattering events [ES] at Super--Kamiokande  
is defined as
\begin{eqnarray} 
&& {\rm [ES]}  \equiv \frac{N_{ES}}{N_{ES}^{SSM}} 
=    \nonumber\\ &&
\frac{1}{N_{\nu e}}\int_{E_{th}} dE_e' R(E_e, E_e') \int dE_{\nu}
F(E_{\nu})
\sigma_{e}(E_{\nu},E_e')
[P_{ee}(E_{\nu}, \theta_Z) + 
r(E_{\nu} ,E_e')(1 - P_{ee}(E_{\nu}, \theta_Z)), 
\label{ESrate}
\end{eqnarray}
where $R(E_e, E_e')$ is the energy resolution factor, 
$F(E_{\nu})$ is the flux of the electron neutrinos without 
oscillations,  $\sigma_{e}$ is the $\nu_e - e$ elastic scattering 
cross-section and 
$r(E_{\nu} ,E_e') \equiv \sigma_{\mu}/\sigma_{e}$. 
with  $\sigma_{\mu}$ being  the $\nu_{\mu} - e$ elastic scattering 
cross-section \cite{CrSe}. 
$N_{\nu e}$ is the normalization factor which equals 
the integral in (\ref{ESrate})  taken at  $P = 1$. 

The reduced rate of the charged current events [CC] at SNO 
is obtained from 
\begin{equation}
{\rm [CC]} \equiv \frac{N_{CC}}{N_{CC}^{SSM}} = 
\frac{1}{N_{\nu d}}\int_{E_{th}} dE_e' R(E_e, E_e') \int dE_{\nu}
F(E_{\nu}) 
\sigma_{CC}(E_{\nu},E_e')
P_{ee}(E_{\nu}, \theta_Z),  
\label{CCrate}
\end{equation}
where $\sigma_{CC}$ is the CC neutrino--deuteron cross-section 
\cite{crossdeu}, and 
$N_{\nu d}$ is the normalization factor  which equals the integral in
(\ref{CCrate})  taken at  $P = 1$. 

\begin{figure}
\centerline{\psfig{figure=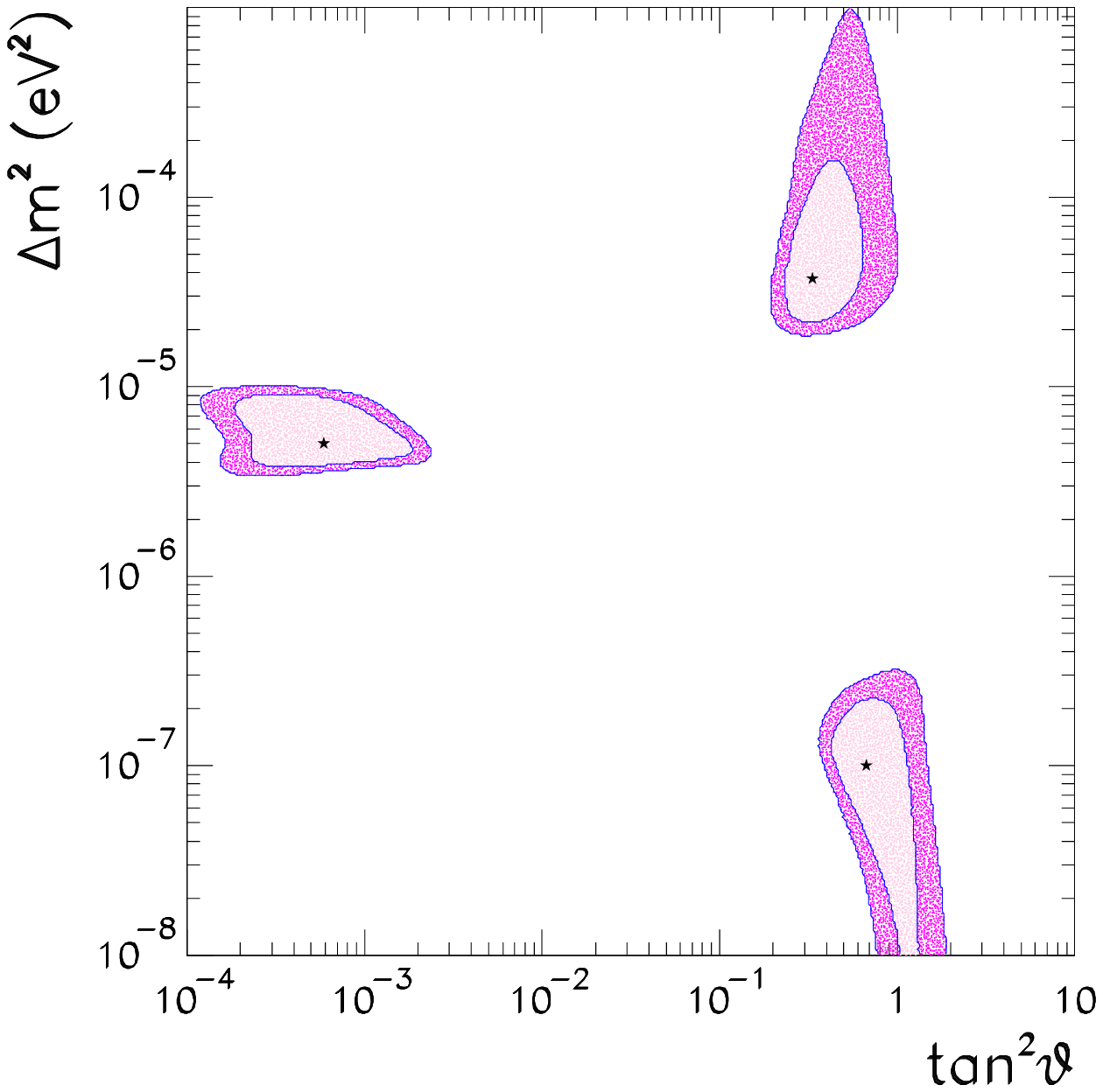,width=5.0in,angle=0}}
\tightenlines  
\caption{Regions of solutions of the solar neutrino problem in the 
$\Delta m^2 - \tan^2 \theta$ plane.  
Dots correspond to the best fit points for each type of the solution. 
Contours show 90 \% CL regions (light) and 99 \% CL regions (dark) 
found with respect to local minima for each region.}
\label{fit}
\end{figure}

The results presented in this section have been obtained for 
oscillation parameters in the presently allowed regions of solutions of 
the solar neutrino problem.  
In Fig.~\ref{fit} we show the  results of the global fit to the data 
which include 
(i) the SuperKamiokande (SK)  data after  1117 days of operation 
(total number of events and day and night energy spectra), 
(ii) Gallex, GNO and  SAGE data, 
(iii) the Homestake data. Shown are the best fit points 
for each solution as well as 90\% and 99\% CL  regions found from the 
local minima in each allowed region. We show the solutions
only in the range  $\Delta m^2 > 10^{-8}$ eV$^2$ for which Earth--matter
effects  on the boron neutrinos can be substantial. The fit includes the
latest 
standard solar model fluxes, BP00 model~\cite{BP00}. For details of the 
statistical analysis applied to the different observables we refer 
to Ref.~\cite{nu2000}. 

\begin{figure}[!t]
\centerline{\psfig{figure=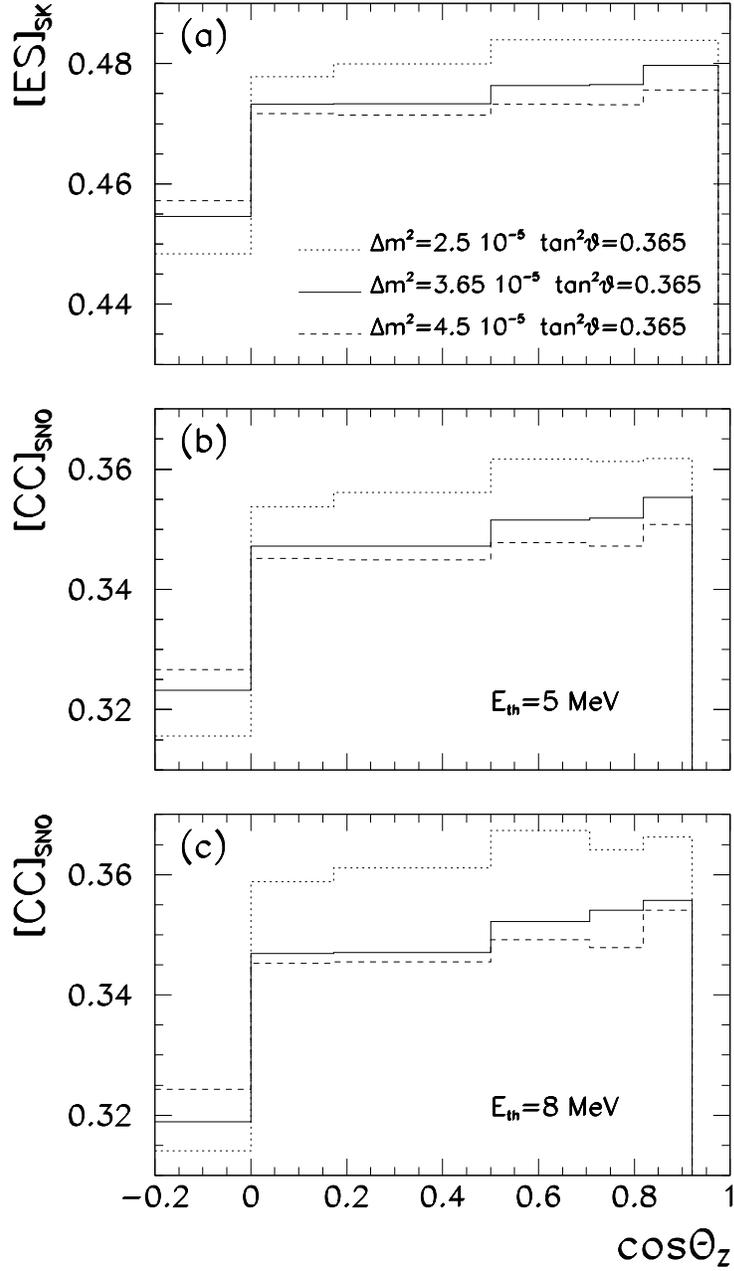,width=5.0in,angle=0}}
\tightenlines  
\caption{The binned zenith angle dependence of (a) [ES] event 
rates at SK and  (b) the $\nu d$ CC -  event rate at SNO above 
at $E_{th} = 5$ MeV and (c) $E_{th} = 8$ MeV  in the LMA region. 
The histograms correspond to different values 
of the oscillation parameters as labeled in the figure. 
The first bin corresponds to the day rate. The event rates have been
computed with a boron flux normalization chosen to fit the measured rate
at SK.} 
\label{bin-lma}
\end{figure}

\begin{figure}[!t]
\centerline{\psfig{figure=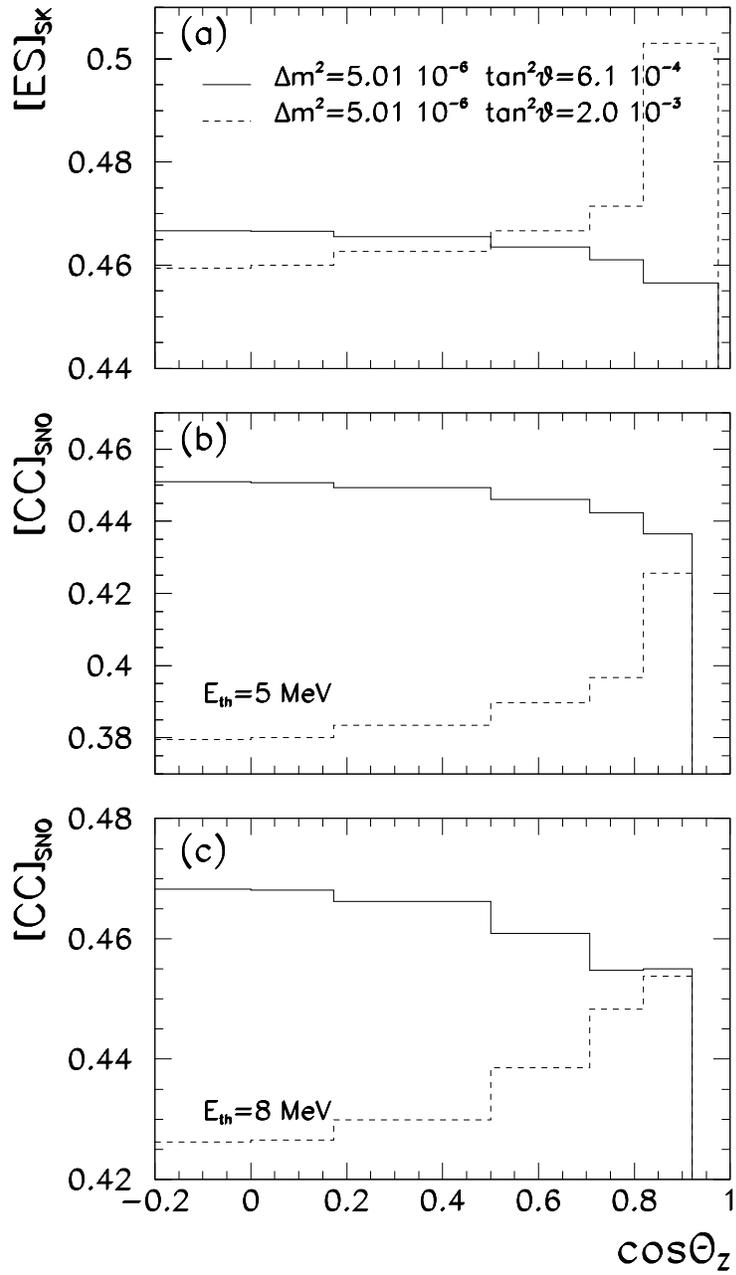,width=5.0in,angle=0}}
\tightenlines  
\caption{Same as Fig.~\protect{\ref{bin-lma}} for the SMA solution. }
\label{bin-sma}
\end{figure}

\begin{figure}[!t]
\centerline{\psfig{figure=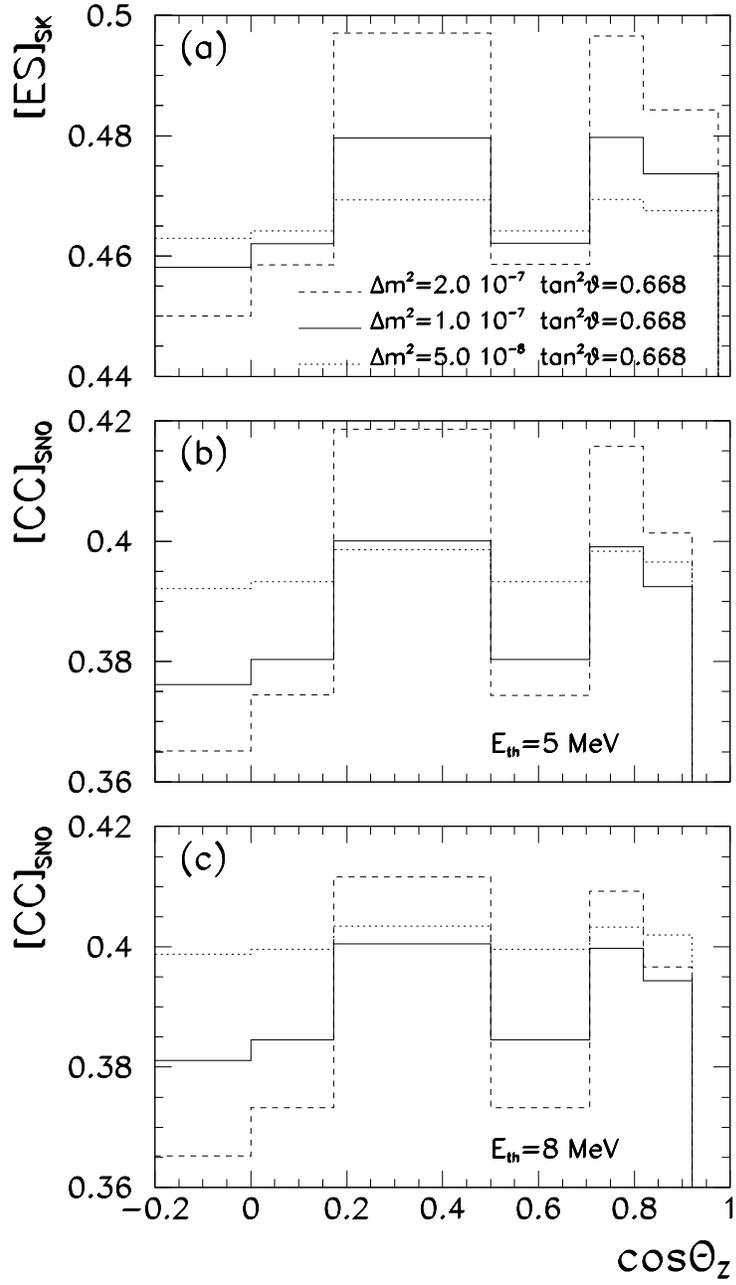,width=5.0in,angle=0}}
\tightenlines  
\caption{Same as Fig.~\protect{\ref{bin-lma}} for the LOW solution. }
\label{bin-low}
\end{figure}

The results for the event rates presented
in Figs.~\ref{reg-lma}--\ref{bin-low} have been normalized to 
the SK total rate. That is, for each set  of the oscillation 
parameters, expressions in Eqs.~(\ref{ESrate}) and~(\ref{CCrate}) have
been  multiplied by a boron flux normalization 
factor $f_{SK}$ in order get value of [ES] as it is  measured  
at SK.   

\subsection{LMA}
\label{sec:lma}

In the LMA region, the adiabaticity condition is satisfied and 
we can safely put $P_c=0$. Consequently Eq.~(\ref{probtot}) simplifies to:
\begin{equation}
P_{ee} = P_D -  \cos2\theta_S \freg, 
\label{Plma}
\end{equation}
where 
\begin{equation}
P_D =  \frac{1}{2} -  \frac{1}{2} \cos2\theta_S \cos2\theta ~.               
\label{Plma2}   
\end{equation}
Moreover, in the low $\Delta m^2$ part of the LMA region we can take 
$\cos2\theta_S \approx -1$,  so that 
\begin{equation}
P_{ee} = \sin^2 \theta + \freg ,  ~~~~~(\Delta m^2 \ll 10^{-4} 
{\rm eV^2}). 
\label{Plma3}
\end{equation}

\begin{figure}[!t]
\centerline{\psfig{figure=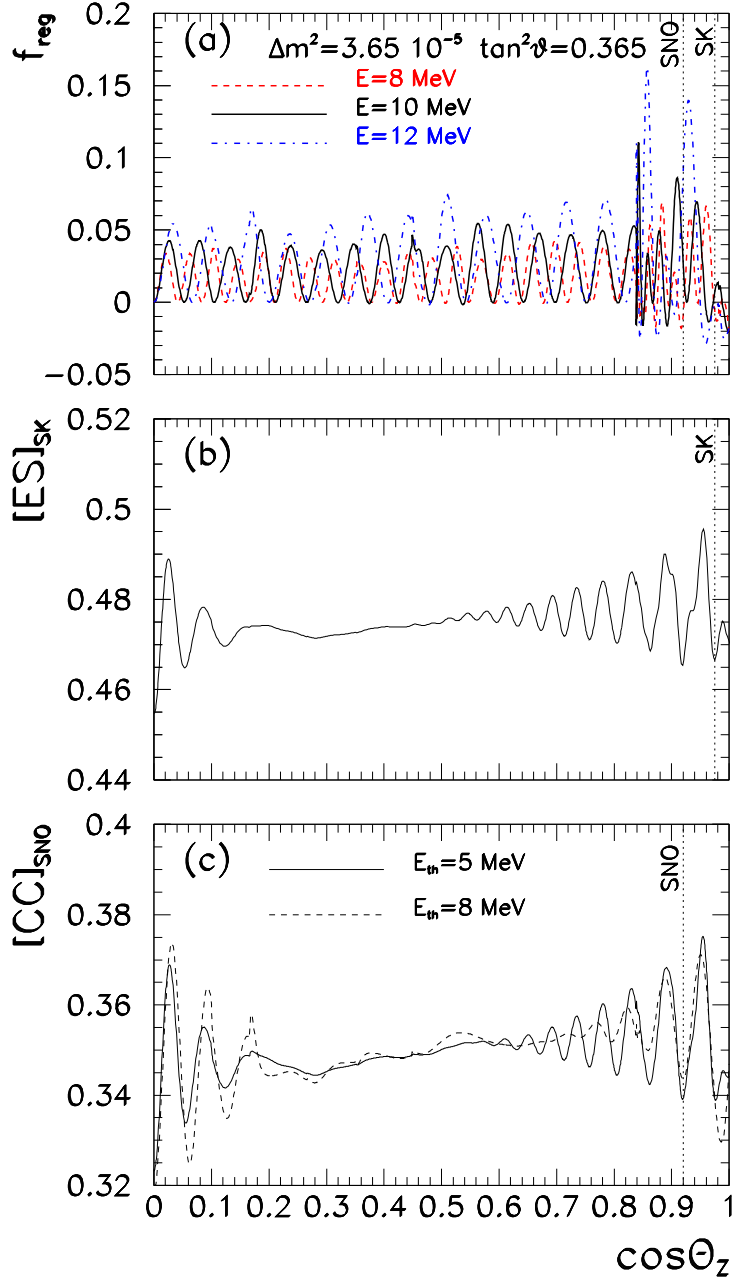,width=5in,angle=0}}
\tightenlines  
\caption{The zenith angle distributions 
for the best fit point ($\Delta m^2 = 3.25 \times 10^{-5}$  eV$^2$, 
$\tan^2 \theta = 0.365 $) in the LMA solution.
(a) The dependence of the regeneration factor on $\cos \theta_Z$ 
for different neutrino energies as labeled in the figure.  
Vertical dashed lines indicate the maximal values of $\cos \theta_Z$ 
which can be realized at SK and SNO. 
(b) The zenith angle dependence of the 
$\nu e$ event rate at Super--Kamiokande above the energy $E_{th} = 5.5$ MeV. 
(c) The zenith angle dependence of the
$\nu d$  charged current event rate at SNO for 
two different energy thresholds (5 and 8 MeV ).}
\label{reg-lma}
\end{figure}

\begin{figure}[!t]
\centerline{\psfig{figure=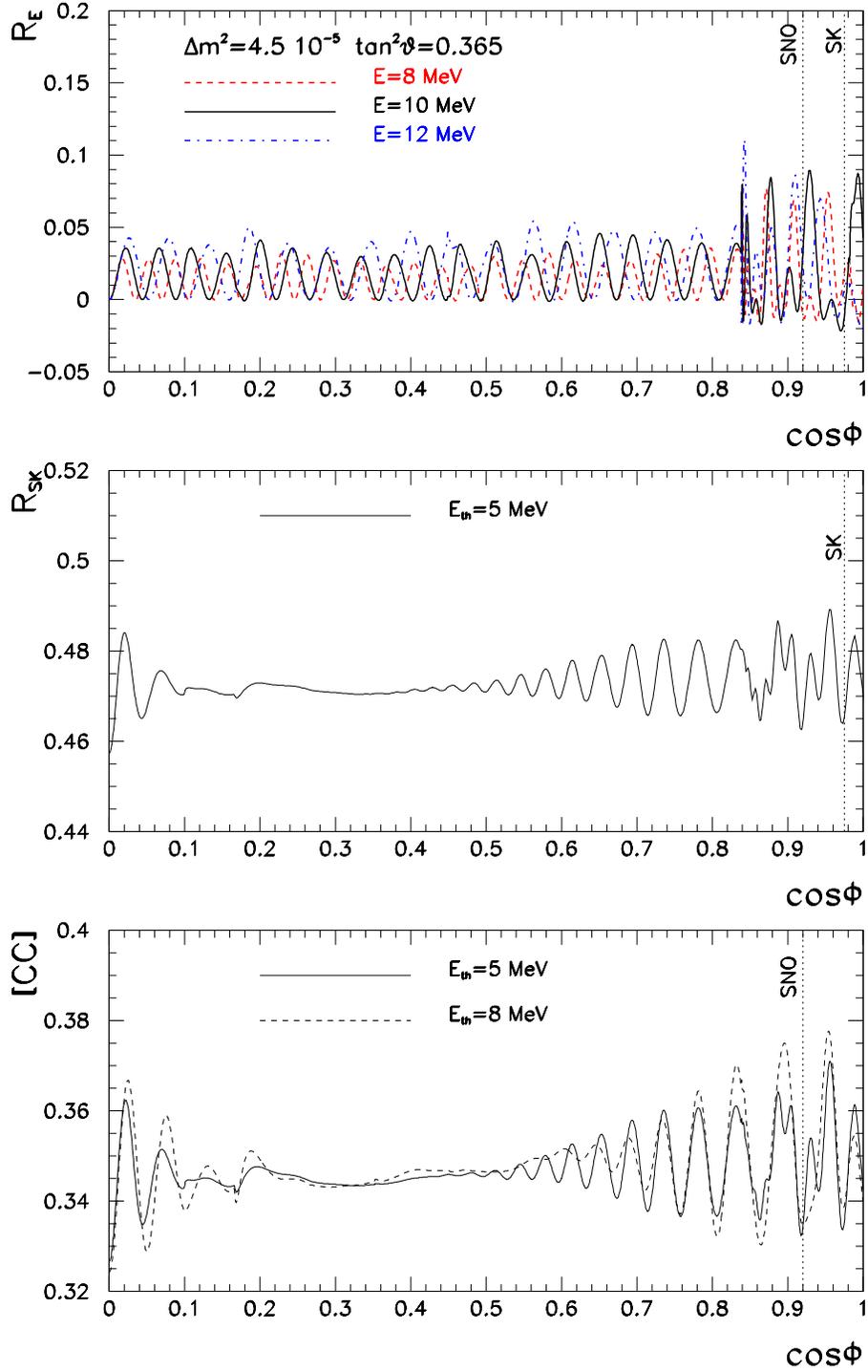,width=5.0in,angle=0}}
\tightenlines  
\caption{Same as Fig.~\protect{\ref{reg-lma}} 
for a larger $\Delta m^2$
 ( $\Delta m^2 = 4.5 \times 10^{-5}$  eV$^2$).}
\label{reg-lma-up}
\end{figure}

Let us discuss first the behaviour of the regeneration factor. 
In Figs.~\ref{reg-lma}a  and~\ref{reg-lma-up}a  we show the zenith angle 
dependence of the $\freg$ for different values of neutrino 
energies. In the interval $\cos \theta_Z = 0 - 0.837$ the dependence has a
quasi-periodical 
shape. The amplitude of oscillation slightly changes  with 
$\cos \theta_Z$ as a consequence of  the adiabaticity violation 
(arising from the small jumps of the density).    
At $\cos \theta_Z >  0.837$ the dependence is more irregular due to 
the Earth core effect. All these features, as well as 
the dependence of the regeneration factor  
on the neutrino parameters (energy, mixing angle, and $\Delta m^2$) 
can be immediately understood from our analytical consideration as
we discuss next. 

In the LMA region of oscillation parameters the Earth--matter effects can
be considered  as 
neutrino oscillations in matter with slowly changing  density. 
Moreover,  in this region $\eta > 3$    
(for the present best fit point (see Fig.~\ref{reg-lma})  
and $E\sim 10$ MeV we get $\eta = 12$). 
This means that oscillations in the Earth with parameters from the LMA
region proceed in the {\it vacuum dominated} regime, when the matter
effect gives relatively small corrections. In particular, the oscillation 
length is close to the vacuum oscillation length: 
\begin{equation}
l_m \approx  l_{\nu}\left[1 + \frac{2 \cos2 \theta}{\eta}\right] 
\label{lengtht}
\end{equation}
which holds to first order in $1/\eta$ 
and presents a rather weak dependence on the mixing angle. 
For the best fit  point oscillation parameters 
and $\rho Y_e \sim 2$ g/cc we get 
\begin{equation}
l_m \approx  10^3~{\rm km} \left(\frac{E}{10 {\rm MeV}}\right)~. 
\label{lengthest}
\end{equation}       
So, the oscillation length is much shorter than the diameter of the Earth
for all the relevant oscillation parameters.

The expression for the mixing angle can be approximated as  
\begin{equation}
\tan 2\theta_m \approx  \tan 2\theta \left[1 + 
\frac{1}{\cos2 \theta \eta}\right]\, ,
\label{angle}
\end{equation}
and the oscillation phase acquired by neutrinos crossing the Earth equals 
\begin{equation}
\phi \equiv  \frac{2 \pi d(\theta_Z)}{l_m(\theta_Z)} \approx 
\frac{2 \pi D \cos \theta_Z}{l_{\nu}}, 
\label{phase}
\end{equation}
where $D = 1.3 \times 10^{4}$ km  is the diameter of the Earth. 
That is, the phase increases linearly with $\cos \theta_Z$, 
and therefore, $\freg$ turns out to be an almost periodical function of
$\cos \theta_Z$. Also, the phase is inversely proportional to the energy, 
so that the number of periods increases as 1/E.  
The period of oscillations in the $\cos \theta_Z$ scale equals  
$T(\cos \theta_Z) \approx l_{\nu}/D$.

\begin{figure}[!t]
\centerline{\psfig{figure=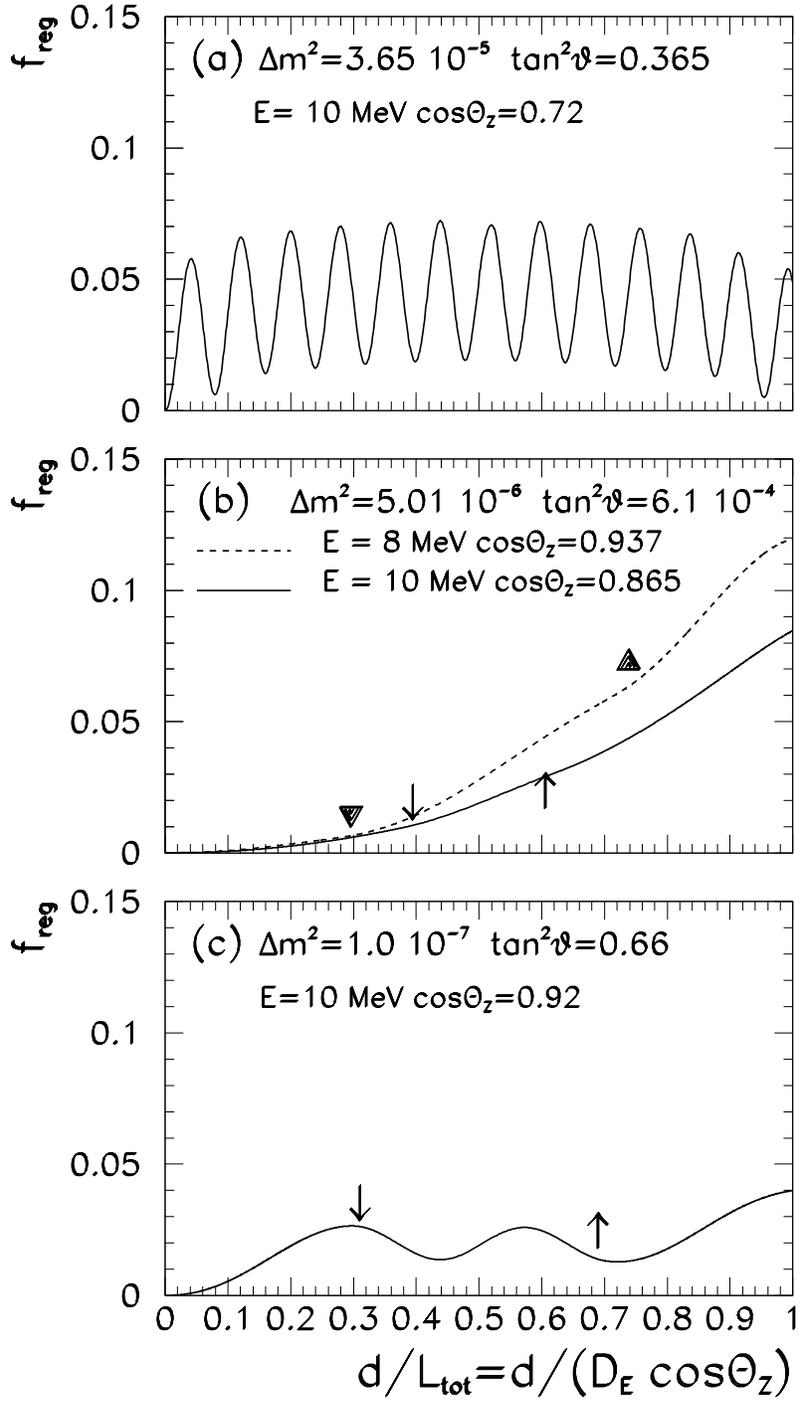,width=4.5in,angle=0}}
\tightenlines
\caption{The dependence of the regeneration factor on distance 
in units of the total length  of a trajectory for a given  
value of $\cos \theta_Z$ and for the oscillation parameters indicated in
the pannels. The dependence is shown for 
(a) LMA solution, (b) SMA solution for two different energies 
(arrow and triangles indicate the point when neutrino enters and leave the
core),  (c) LOW solution.}
\label{evol}
\end{figure}

In fig.~\ref{evol} we show the dependence of $\freg$ on the distance 
travelled by the neutrino inside the Earth for some trajectories and 
for best fit values of the oscillation parameters. For LMA 
the oscillation length is much smaller than  the typical scale of the 
density change: 
\begin{equation}
l_{\rho} \equiv \rho \left(\frac{d\rho}{dx} \right)^{-1}  \sim \frac{D}{2}
\label{gradient}
\end{equation}
both for the mantle and for the core: $l_m \ll l_{\rho}$. 
Since the mixing angle is large this leads to  good adiabaticity 
(especially far from the resonance). Small density jumps 
can be treated then as small perturbations. As a result, the average 
$\freg$, $\overline{\freg}$, follows the density change. 
Both $\overline{\freg}$ and the amplitude of oscillations increase towards 
the center of the trajectory where the density 
is maximal. Since the profile is symmetric (with respect to the middle 
point of the trajectory), the dependence of $\overline{\freg}$ 
and 
$A_f$ on distance is also symmetric. At the detector $\overline{\freg}$ 
and $A_f$ are determined by the surface density ($\rho \sim 3.2$ g/cc). 
Therefore in the adiabatic approximation the amplitude of the oscillatory 
behavior of $\freg$ should not depend on $\cos
\theta_Z$. The variations of the amplitude  $A_f$ with $\cos \theta_Z$ 
seen in Fig.~\ref{evol} are produced by the small density jumps 
which violate the adiabaticity. 

As follows from Eq.~(\ref{depthA}), the amplitude of oscillations takes the 
form
\begin{equation}
A_f \approx  \frac{\sin^2 2\theta}{\eta_{s}} k_{na}(\theta_Z) 
\approx \frac{E \rho_{s}}{\Delta m^2}~,
\label{depth2}
\end{equation}
where $\eta_{s}$ and  $\rho_{s}$ are  the value of parameter 
$\eta$ and the density at the
surface of the Earth and 
$k_{na} \sim 1$ is the parameter which describes small effects of the
adiabaticity violation. 
Thus,  the amplitude  increases with energy, as can be also seen from 
Figs.~\ref{reg-lma}a and~\ref{reg-lma-up}a,  
and it is inversely proportional to $\Delta m^2$. 
The amplitude is  proportional to $\sin^2 2\theta$, 
however variations of this  parameter in the LMA region 
produces small changes specially in the near maximal mixing region. 

Combining the results of Eqs.~(\ref{phase}) and (\ref{depth2}) we get 
the zenith angle dependence of the regeneration factor in the 
mantle region 
\begin{equation}
\freg \approx 
\frac{\sin^2 2\theta}{\eta_{s}} k_{na}(\theta_Z) 
\sin^2 \frac{\pi D \cos \theta_Z}{l_{\nu}}~. 
\label{zenmantl}
\end{equation}

The zenith angle dependence  becomes more complicated in the region 
$\cos \theta_Z > 0.837$, when neutrino trajectories cross 
the core of the Earth \cite{pet,akh,CP,AkS,param_above}
. This is related to the appearance of the 
parametric (enhancement and suppression) effects on the top of the
resonance enhancement in the core. 
Notice that for core-crossing trajectories the regeneration factor can be
negative. In contrast, in  the mantle region it is always
positive as can be seen from Eq.~(\ref{zenmantl}) (see also Appendix). 

Finally let us point out that the results shown in 
Figs.~\ref{reg-lma}--~\ref{reg-low} for the regeneration factor are 
independent of the experimental set up.
However the range of zenith angles covered by a given 
detector depends on its latitude: for SK $\cos \theta_Z^{max} = 0.975$,  
whereas for the SNO  $\cos \theta_Z^{max} = 0.92$. For the sake of 
clarity in Figs.~\ref{reg-lma}--~\ref{reg-low} we mark those limiting 
zenith angles. 

Let us now consider the zenith angle dependence of the event rates. 
We have calculated the rates above a given energy threshold using 
Eqs.~(\ref{ESrate}) and~(\ref{CCrate}). The calculation of rates involves 
folding of the survival probability with neutrino cross-section, 
the flux of neutrinos and integration over the energy above threshold.
Therefore results are  different for SK and SNO. 

In the lower part of the LMA region the survival probability 
is simply given by the sum of $\sin^2 \theta$ and the regeneration
factor as seen in Eq.~(\ref{Plma3}). Consequently, the [CC] rate  
can be written as 
\begin{equation}
{\rm [CC]} \approx \sin^2 \theta + \langle \freg(\theta_Z)
\rangle_{CC}
\label{CC_rate_av}~,  
\end{equation}
where the averaged regeneration factor is given by the equation 
(\ref{CCrate}) with $P_{ee}$ substituted by $\freg$. 

Similarly, the [ES] rate can be expressed as 
\begin{equation}
{\rm [ES]} \approx \sin^2 \theta +  r \cos^2 \theta + 
(1 - r) \langle \freg(\theta_Z) \rangle_{ES}~, 
\label{ES_rate_av}
\end{equation}
where $\langle \freg(\theta_Z) \rangle_{ES}$ is given by 
Eq.~(\ref{ESrate}) with $P_{ee}$ substituted by $\freg$. Notice that 
the interactions of $\nu_{\mu}$ and $\nu_{\tau}$   
enhance the average [ES] rate in comparison with [CC] rate 
giving an additional term  $r \cos^2 \theta$, and they 
suppress the regeneration term by a factor $(1 - r)$. 

In Figs.~\ref{reg-lma}b and~\ref{reg-lma-up}b 
we plot the rate [ES] at SK as a function 
of $\cos \theta_Z$. The dependence of the reduced [CC] rate at SNO on 
$\cos \theta_Z$ for two different 
thresholds is shown in Figs.~\ref{reg-lma}c and~\ref{reg-lma-up}c.   
(Similar figures have been obtained in \cite{BW,BK}). 
As seen in the figures, the rates are oscillating functions of 
$\cos \theta_Z$, however in contrast with the case of the regeneration 
factors,  the  amplitude of these oscillations changes with the zenith angle
significantly. 
The amplitude is maximal 
at small $\cos \theta_Z$ ($\cos \theta_Z = 0 - 0.2$), it decreases 
with $\cos \theta_Z$ till $\cos \theta_Z \sim  0.5$,  and then it
increases again. 
The first maximum is achieved already at $\cos \theta_Z = 0.02 - 0.03$ 
which corresponds to the distances 250 - 400 km. This large 
matter effect on small distances is related to the fact that 
the initial state  is not a flavor state but an incoherent 
admixture of the mass eigenstates $\nu_1$ and $\nu_2$ \cite{akhm00}. 
Also the period of oscillations changes: it is about 
0.07 for small $\cos \theta_Z$ and it decreases down to 0.05 for 
$\cos \theta_Z > 0.6$. 
This important feature is related to the integration over the 
neutrino energy  
and it can be explained  as the {\it synchronization} effect 
of the ``waves" (oscillatory curves)  $\freg (\theta_Z)$  corresponding to
different neutrino energies as we discuss next. 

Let us  notice that the boron neutrino flux is maximal at $E = 8$ MeV 
and it decreases with energy, while both the cross-section and the 
regeneration factor increase with $E$. As a result, the main
contribution to the integrated regeneration factor,    
$\langle \freg(\theta_Z) \rangle$,  comes from a rather narrow interval
of energies:  $E = 9 - 13$ MeV, so that $\Delta E / E \sim 1/3$. 
For the central value of the energy $E = 11$ MeV and oscillation parameters 
in  the best fit point (see Fig.~\ref{reg-lma})
the oscillation length is about $10^3$ km. This distance corresponds to 
$\cos \theta_Z \sim 0.07$. That is,  for neutrinos propagating at  
$\cos \theta_Z \sim 0.07$  the oscillation phase will be 
$\phi \approx 2 \pi$ and $freg$ will be at a minimum. 
In general, we get from Eq.~(\ref{phase})
\begin{equation}
\phi = 14 \pi  \cos \theta_Z 
\left(\frac{\Delta m^2}{3 \times 10^{-5} {\rm eV}^2}\right) 
\left(\frac{10 {\rm MeV}}{E}\right)~. 
\label{phase_num}
\end{equation}
Therefore,  the phase difference $\Delta \phi$ for  neutrinos 
which cross the Earth in the direction $\cos \theta_Z$ and
differ in energy by $\Delta E$ is
\begin{equation}
\Delta \phi \approx  \phi \frac{\Delta E}{E} 
\approx 2 \pi \frac{\cos \theta_Z}{0.07}\frac{\Delta E}{E}~.
\label{phase_dif}
\end{equation}
If $\Delta \phi \sim  2\pi$ one would expect strong averaging effect.  
From this condition and using Eq.~(\ref{phase_dif}) with 
$\Delta E/E = 1/3$ we find that strong averaging should appear at 
$\cos \theta_Z \approx 0.2$ in agreement  with the results of the
numerical 
calculations,  as seen in Figs.~\ref{reg-lma}b and ~\ref{reg-lma}c.
For large $\cos \theta_Z$, the waves are again synchronized,   
so that a constructive interference of the  waves leads to the
restoration of the oscillatory behavior. 

From Eq.(\ref{phase_dif}) we also see that 
the periods of constructive and destructive interference 
become shorter with the increase of 
$\Delta m^2$ (see Fig.~\ref{reg-lma-up}). Also  
with the increase of the energy threshold the relevant energy interval 
$\Delta E$ becomes narrower and, correspondingly, the period of
synchronization
becomes longer  (in $\cos \theta_Z$ scale) as can be seen in 
Fig.~\ref{reg-lma}c and~\ref{reg-lma-up}c. 

\subsection{SMA}
\label{sec:sma}    

In the SMA region we have $\cos2\theta_S \approx -1$. Furthermore,   
the adiabaticity is broken inside the Sun:  $P_c \neq 0$,  so that  
\begin{equation}
P_{ee} = P_D + (1 - 2 P_c) \cdot  \freg \, .  
\label{Psma1}   
\end{equation}
Notice that since in the SMA region $\cos 2\theta\sim 1$,  one has  
$P_D  \approx P_c$. SK data imply that $P_c \approx P_D = 
0.35$--$0.65$, therefore in the SMA region the regeneration
effect is substantially suppressed by the adiabaticity factor: $|1 - 2
P_c| < 0.3$. For $P_c > 1/2$ the regeneration effect becomes negative.    

\begin{figure}[!t]
\centerline{\psfig{figure=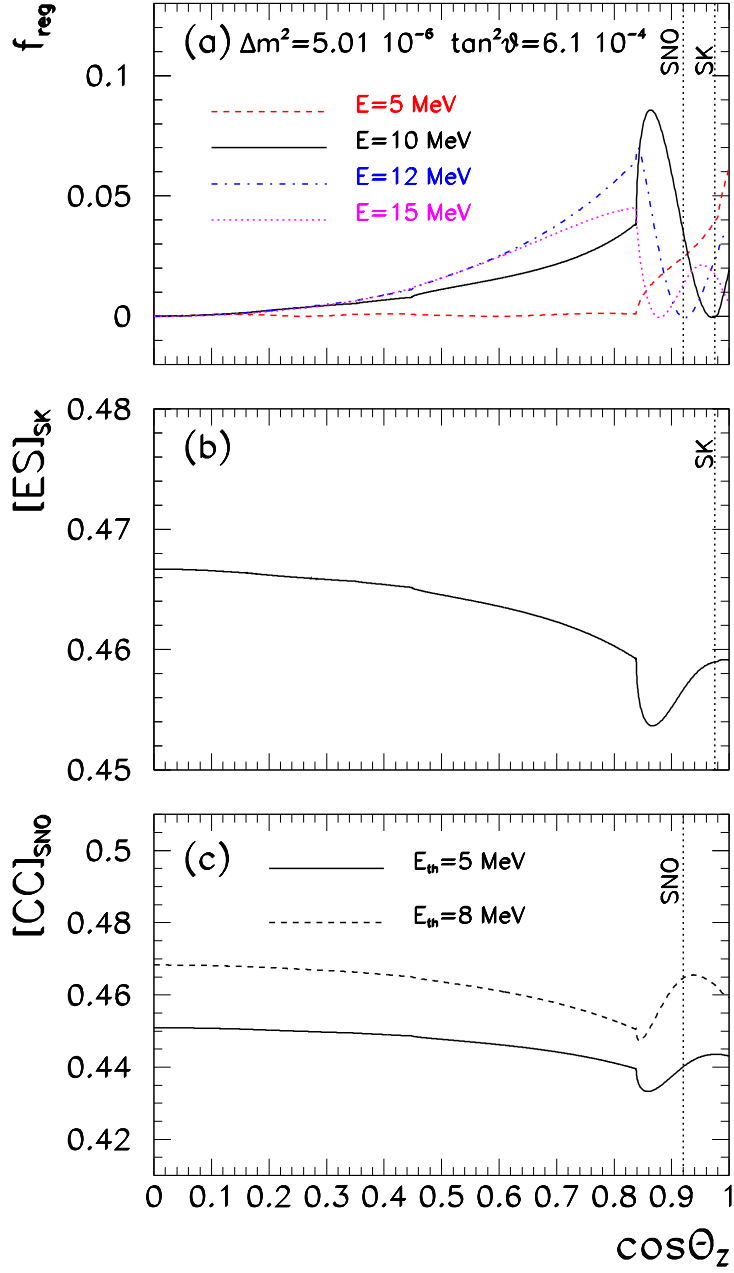,width=5.0in,angle=0}}
\tightenlines  
\caption{Same as Fig.~\protect{\ref{reg-lma}} 
for the best fit point 
( $\Delta m^2 = 5.01 \times 10^{-6}$  eV$^2$, 
$\tan^2 \theta = 6.1 \times 10^{-4} $ ) in the SMA solution.}
\label{reg-sma1}
\end{figure}

In Fig.~\ref{reg-sma1}a we show the zenith angle dependence of the
regeneration factor for different values of neutrino energy and 
for oscillation parameters  corresponding to the best fit point in the SMA
region. 
The behaviour of the curves can be understood taking into account that 
in the SMA region 
\begin{equation}
\eta = 0.3 - 3~, 
\label{sma_eta}
\end{equation}
so the SMA solution region covers the resonance range for the Earth
matter densities. For the best fit point we have 
\begin{equation}
l_{\nu} \approx l_0 \approx  (4 - 8) \times 10^3 ~ {\rm km}~.  
\label{phase_sma}
\end{equation}

For the mantle region ($\cos\theta_Z < 0.837$) we find that:\\
1. Far from the resonance, $\eta > 1$,  
the oscillation length,  
\begin{equation}
l_m \approx \frac{l_{\nu}\eta}{|1 - \eta|}
\label{length-far}
\end{equation}
is smaller than the Earth diameter. However the amplitude of oscillations is
strongly suppressed (see the line which corresponds to the 
$E = 5$ MeV in fig.~\ref{reg-sma1}a)\\ 
2. In the resonance, $\eta=1$, 
the amplitude of oscillations is strongly enhanced, but  the oscillation
length
\begin{equation} 
l_m \approx \frac{l_{\nu}}{\sin^2 2\theta} \approx 10^5~ {\rm km}~,
\label{length-res}   
\end{equation}  
is much larger  that the Earth diameter. As a result, the oscillation
effect is small (the line for $E = 12$ MeV in fig.~\ref{reg-sma1}).  
In the small phase limit the regeneration factor equals 
\begin{equation}
\freg \approx
\sin^2 2\theta \cdot 
 \frac{\pi^2 D^2 \cos^2 \theta_Z}{l_{\nu} l_0}~.   
\label{f-sma2}
\end{equation}
It increases quadratically with 
$\cos \theta_Z$. \\
3. The line for $E = 15$ MeV represents an intermediate case when the
phase of oscillations is about $\pi$ for $\cos \theta_Z \approx 0.8$. 
In this case the regeneration factor equals:
\begin{equation}
\freg \approx
4 \sin^2 2\theta \cdot
 \frac{D^2 \cos^2 \theta_Z}{l_{\nu} l_0}~
\approx 4 \sin^2 2\theta \cdot \frac{D^2 \cos^2 \theta_Z}{ l_0^2}, 
\label{f-sma3}
\end{equation}
where in the second equality we have taken into account  that  $\eta \sim
1$. 
Since $l_0 \approx 0.4 D$ for large $\cos \theta_Z$ trajectories in the
mantle, the mixing enhancement factor,  $4 (D/l_0)^2$, can be as large as 
25, so that $\freg \approx 0.04$ for $\sin^2 2\theta = 2.4 \times 10^{-3}$  
and $\cos \theta_Z = 0.8$. 

Clearly, the adiabaticity is strongly broken near the resonance: 
$l_m \gg l_{\rho}$. But far from the resonance: $l_m \sim  
l_{\rho} \sim D$,  the violation of the adiabaticity is moderate, so one 
can describe its effect as oscillations in a narrow layer with some 
effective density.

For $\cos \theta_Z > 0.837$  neutrinos cross the core of the Earth. 
For the core densities the resonance energies are in the range 
$E = 3 - 5$ MeV ($\Delta m^2 \sim 5 \times 10^{-6}$ eV$^2$).  
For energies between $E_R(core)$ and  $E_R(mantle)$, the parametric 
enhancement of oscillations takes place leading to the appearance of the
parametric peak in the $\cos \theta_Z$ distribution of $\freg$. 
For SMA the realization of the parametric resonance corresponds to a
mixing angle in mantle smaller than  maximal
mixing, $2\theta_{mantle}<\pi/2 $, 
while in the core,  $2\theta_{core} >  \pi/2$. 
At the peak the relation  
between the oscillation phases in the mantle $\phi_{mantle}$ and in the
core $\phi_{core}$ corresponds to the general
condition for the parametric resonance \cite{akh,CP,AkS}
\begin{equation}
X_3 \equiv s_m c_c \cos 2 \theta_{mantle} + s_c c_m\cos 2 \theta_{core}
\approx 0 \,,
\label{parres}
\end{equation}
where $s_m \equiv \sin (\phi_{mantle}/2)$, 
$c_c \equiv \cos (\phi_{mantle}/2)$.  
For the best fit point  $\tan^2 \theta  = 6.1 \times 10^{-4}$, 
$\Delta m^2 = 5\times 10^{-6}$ eV$^2$ and $E = 10$ MeV 
we find that the maximum
of the parametric peak occurs at $\cos \theta_Z = 0.863$. 
The   phases in the core and in the mantle are 
$\phi_{mantle} =  0.49 \pi$,    $\phi_{core} = 0.49 \pi$, and  
the effective mixing angles 
are  $2\theta_{mantle} =0.188$ and $\pi - 2\theta_{core} = 0.113$. 
Then we get $X_3 \sim 0.045 \ll 1$ . 

As illustration, in Fig.~\ref{evol}b we show  the dependence of the 
regeneration factor on 
the distance for the parameters which correspond to the maxima of the
parametric peaks. 

\begin{figure}[!t]
\centerline{\psfig{figure=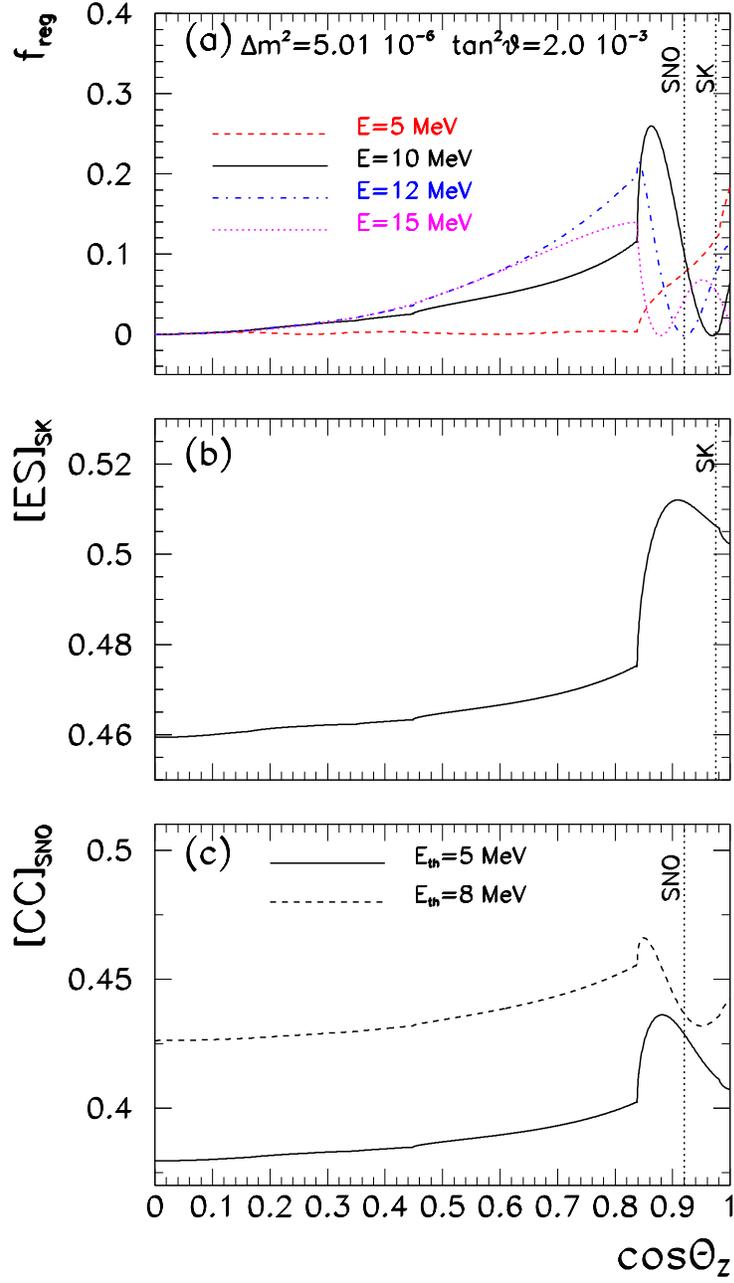,width=5.0in,angle=0}}
\tightenlines  
\caption{Same as  Fig.~\protect{\ref{reg-sma1}} 
for a larger mixing 
($\tan^2 \theta = 2.0 \times 10^{-3} $ ) in the SMA solution.}
\label{reg-sma2}
\end{figure}

In Figs.~\ref{reg-sma1}a and~\ref{reg-sma2}a  we show  the 
dependence of the 
regeneration factor on the mixing angle. Both the oscillation length and 
the resonance condition depend on $\sin^2 2\theta$ weakly, so that with 
a good precision 
$\freg \propto \sin^2 2\theta$.  Notice that this proportionality 
holds also for core crossing trajectories. 

In Figs.~\ref{reg-sma1} and~\ref{reg-sma2} (panels b and c) we show 
the zenith angle dependence of the event rates at SK and SNO. 
The integration over energy basically reproduces the dependence of 
$\freg$ on $\cos \theta_Z$ for $E \sim 10$ MeV since this energy 
gives the dominant contribution. 
The [CC] event 
rate at SNO  can be writen as
\begin{equation}
{\rm [CC]} \approx  
\langle P_D \rangle_{CC} + (1 - 2 \bar{P}_c)   
\langle \freg(\theta_Z) \rangle_{CC}   
\label{CC-sma}~,
\end{equation}
where $\bar{P}_c$ is an effective jump probability for the contributing  
energies.  (Notice that ${P}_c$ changes  slower with $E$ than 
$\freg$ does.) 

For $\tan^2 \theta \sim   6 \times 10^{-4}$ (see Fig.~\ref{reg-sma1})
the adiabaticity breaking 
is strong, $\bar{P}_c > 1/2$ and the negative adiabatic factor leads to
the suppression of the rate due to Earth--matter effect. On the other hand, 
for $\tan^2 \theta \sim   2.0 \times 10^{-3}$ (see fig.~\ref{reg-sma2}), 
the adiabaticity 
breaking is weaker,  $\bar{P}_c < 1/2$, and the Earth--matter effect 
is positive.  
As follows from the figures, the adiabatic factor 
suppresses substantially the regeneration effect: 
\begin{equation}
|1 - 2 \bar{P}_c|
\frac{\langle \freg(\theta_Z) \rangle_{CC}}{\freg} \lsim 0.2 - 0.3~.
\label{CCad-sma}
\end{equation}

The [CC] rate depends weakly on 
the energy threshold in the mantle region but it strongly decreases with 
$E_{th}$ in the core region.  
The parametric peak is wider (in energy scale) than the resonance peak 
in the mantle and therefore  
the integration over the energy leads to a stronger 
decrease of the regeneration effect in the mantle region than in the core
region. 
Also the parametric peak  is situated at the low energy part of the spectrum: 
(5 - 10 MeV). As a consequence,  
the core peak decreases substantially with the increase of the threshold 
from 5 to 8 MeV.  

\subsection{LOW}
\label{sec:low}    
In the  LOW solution region  one has 
$\cos2\theta_S \approx -1$, and moreover, to good approximation we can
take $P_c \approx 0$, so that 
\begin{equation}
P_{ee} \approx \sin^2 \theta + \freg. 
\label{pee_low}
\end{equation}

\begin{figure}[!t]
\centerline{\psfig{figure=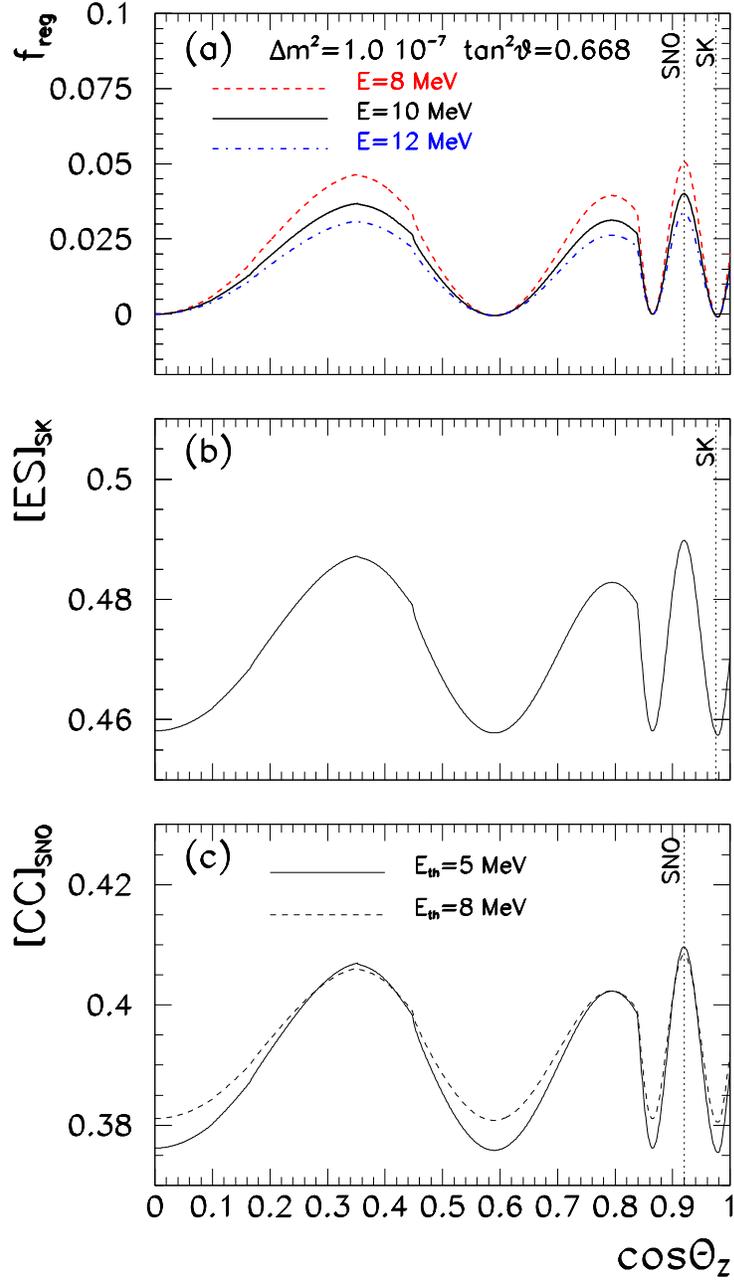,width=5.0in,angle=0}}
\tightenlines  
\caption{Same as  Fig.~\protect{\ref{reg-lma}}  for the best fit 
( $\Delta m^2 = 1.0 \times 10^{-7}$  eV$^2$, 
$\tan^2 \theta = 0.668 $ ) in the LOW solution.} 
\label{reg-low}
\end{figure}

The zenith angle dependence of the regeneration factor 
for oscillation parameters in 
the best fit point  and different values of the neutrino energy 
are shown in Fig.~\ref{reg-low}a. The interpretation of the results is
rather straightforward. 
In the LOW region we have $\eta < 0.1$. In particular,  in the best fit
point at E = 10 MeV :  $\eta < 0.03$.  Therefore the oscillations 
proceed in the {\it matter dominated} regime. Thus,  $l_{\nu} \gg l_0$ and 
the oscillation length is determined mainly by the refraction length. 
Notice that in the limit  $\eta \ll 1$ we get from Eq.~(\ref{l-m})
\begin{equation}
l_m \approx l_0 (1 +2 \eta \cos 2\theta)~.  
\label{l-m-low}
\end{equation}
Moreover, in the LOW region the mixing parameter is small: 
$\cos 2\theta < 0.5$ at 99\% C.L.  
(in the best fit point:  $\cos 2\theta \approx 0.2$). 
So, the correction to $l_0$ is further suppressed, for instance,    
$2 \eta \cos 2\theta \sim 0.012$ for 
$E = 10$ MeV. Therefore for a given  trajectory,  the oscillation phase 
is practically independent of  
the neutrino energy and mixing angle 
(see  different curves in Fig.~\ref{reg-low}a).  

The phase of the oscillatory behavior  of $\freg$ with $\cos \theta_Z$  can be
written as 
\begin{equation}
\phi = 
\sqrt{2} G_F \int_0^{D \cos \theta_Z} n_e(x, \cos \theta_Z) dx 
= \frac{2\pi D \cos \theta_Z}{\bar{l}_0 (\cos \theta_Z)},  
\label{phase-low}
\end{equation}
where $\bar{l}_0 (\cos \theta_Z)$ is the average refraction length along
the trajectory determined by $\cos \theta_Z$ and
$n_e=\rho Y_e/m_N$ is the electron number density. 
With increase of $\cos \theta_Z$, the average density increases so,   
$\bar{l}_0$ decreases and the period of oscillations becomes shorter.  
Thud the oscillation phase is determined by $\cos \theta_Z$.  
The first maximum of $\freg$ ($\phi = \pi$)  
is achieved at  
$\cos \theta_Z = 0.35$, the minimum ($\phi = 2 \pi$) lies at 
$\cos \theta_Z = 0.59$ and second maximum occurs at $\cos \theta_Z = 0.8$ 
(see also \cite{BW}).  

The oscillation length is comparable with the scale of
density variations $l_m \approx l_{0} \sim 8 \cdot 10^{3}$ km, where 
$l_{\rho} \sim D/2$ ,  so that the adiabaticity is  broken moderately. 
Since $l_m$ is comparable with the size of the layer, the effect can be
considered as oscillation in the layer of matter with 
some effective constant density. 

The amplitude  of oscillations: 
\begin{equation}
A_f \approx \eta \sin^2 2\theta \approx  \frac{\sqrt{2} 
\Delta m^2 \sin^2 2\theta}{G_F E \bar{n}(\cos \theta_Z)}
\label{depth-low}
\end{equation}
is proportional to $\Delta m^2$  and inversely proportional to $E$.    
It is also inversely proportional to the average density for a given 
trajectory $\bar{n}$ which increases with $\cos \theta_Z$. Correspondingly
the second peak is lower. 
For E= 10 MeV and   oscillation parameters at the best fit point we
get $\eta  = 0.2$ (for the surface density),  and consequently 
according to  Eq.~(\ref{depth-low}) the height 
of the first peak is 
expected to be $\freg = 0.042$ which is slightly larger than the numerical
result  $\freg = 0.035$ shown in  
Fig.~\ref{reg-low}a. The difference of the results is due to the
adiabaticity breaking. 

Combining Eqs.~(\ref{phase-low}) and~(\ref{depth-low}) we get an approximate
expression for the regeneration factor in the mantle region: 
\begin{equation}
\freg \approx  
\frac{\sqrt{2} \sin^2 2\theta}{G_F \bar{n}(\cos \theta_Z)}
\frac{\Delta m^2}{E} 
\sin^2 \left[ \frac{G_F}{\sqrt{2}}\int_0^{D \cos \theta_Z}  
n_e (x, \cos \theta_Z) dx \right]~.
\label{regen-low}
\end{equation}
Notice also that $\freg$ increases with $\sin^2 2\theta$.  

In the core region $\freg$  is enhanced due to the parametric effect. 
Indeed,  we find that for $\cos \theta_Z = 0.92$, which corresponds to 
the position of the third maximum, the phases of oscillations equal:
\begin{equation}
\phi_{mantle} =0.98 \pi, ~~~\phi_{core} = 2.98 \pi~. 
\label{ph-param}     
\end{equation}
That is, they are very close  to $\pi$ and $3\pi$
\begin{equation}
\phi_{mantle} \approx \pi, ~~~\phi_{core} \approx 3\pi~.
\label{ph-param2}
\end{equation}
At this condition in the constant density approximation 
the height of the peak would reach the  value (see Appendix): 
\begin{equation}
\freg \sim \sin^2 (\theta - 4 \theta_{mantle} + 2 \theta_{core}) 
- \sin^2 \theta \approx \sin 2\theta \cdot 
\sin^2 (4 \theta_{mantle} - 2 \theta_{core}),  
\label{peak-param}     
\end{equation}
where $\theta_{mantle}$ and  $\theta_{core}$ are the mixing angles in the 
mantle and in the core correspondingly, and the last equality is valid for 
 $\theta_{mantle} \sim \theta_{core}\sim \pi/2$. 
The peak in Fig.~\ref{reg-low}a is slightly lower than what  
Eq.~(\ref{peak-param}) gives because of the density change. 
Notice that this realization of the parametric enhancement 
corresponds to both mixing angles above the resonance,  
(see \cite{param_above} and Appendix). 
The real time evolution of the neutrino state 
(dependence of the regeneration factor on distance) for the parameters 
corresponding to the peak is shown in fig.~\ref{evol}c.

In Fig.~\ref{reg-low}b and Fig.~\ref{reg-low}c 
we show the zenith angle dependence of [ES] event rates  and [CC] 
event rates at SK and SNO respectively. 
The [CC] rate has the same approximate expression as in 
Eq.~(\ref{CC_rate_av}) for the LMA solution. Since $\freg \propto 1/E$ and the 
phase of oscillations depends very weakly on $E$, we get 
$\langle \freg \rangle_{CC} \approx \freg (\bar{E})$, where 
$\bar{E} \sim (10 - 11)$ MeV is the effective energy of the spectrum. 
Therefore the [CC] distribution reproduces all the features of the 
zenith angle dependence of $\freg$. 
Furthermore,  the regeneration term weakly decreases with the increase 
of the threshold energy from 5  to 8 MeV,  
and it increases with mixing as 
$\langle \freg \rangle_{CC} \propto \sin^2 2\theta$
according to Eq.~(\ref{regen-low}). 

For the [ES] distribution at SK, the results are similar to those for 
[CC] with the only additional feature of a ~25 \% damping effect due to 
the contribution from $\nu_{\mu}/\nu_{\tau}$  scattering  via NC
(see approximate formula in (\ref{ES_rate_av})).  

\section{Identification of the solution. Correlations}
\label{sec:bincor}

\subsection{Binned rates}
\label{subsec:bin}
According to Figs.~\ref{reg-lma}--~\ref{reg-low} the 
LMA,  SMA and LOW solutions present qualitatively different 
zenith angle dependence of the event rates (either [ES] or [CC]).

1) For the LMA solution one expects a significant rate 
already at $\cos \theta_Z \sim 0.03$. 
The first peak is at $\cos \theta_Z = 0.02 - 0.03$.   
The rate has an oscillatory behavior at small,  
$\cos \theta_Z < 0.2$,  and large $\cos \theta_Z > 0.6 - 0.8$,
zenith angles. The border of the oscillatory region at 
high $\theta_Z$ 
depends on the value of the mass difference.
For instance, for $\Delta m^2 \sim 2.5 \times 10^{-5}$ eV$^2$ the second 
oscillatory region starts at $\cos \theta_Z \sim 0.8$ while 
for $\Delta m^2 \sim 4.5 \times 10^{-5}$ eV$^2$
it expands down to $\cos \theta_Z \sim  0.6$. 
The period of oscillations is small,  
$\Delta (\cos \theta_Z) < 0.05 - 0.07$,  and in consequence, it will 
be difficult to detect the oscillatory behaviour due to the relatively 
low statistics. Although we find that for small values of $\Delta m^2$ 
the amplitude of oscillations can reach 15 \% of the average rate. 
When  averaging over wider $\cos \theta_Z$ bins,  the binned
rate depends rather weakly on $\cos \theta_Z$ and  
no significant change of the rate is expected in the core region.

2) For the SMA solution the rate changes slowly and monotonously  in the
mantle region and significant parametric effects can appear in the core
region. For the large $\tan^2 \theta$ the rate increases with 
$\cos \theta_Z$ and  
the parametric enhancement leads to the appearance of the parametric peak 
which gives the main contribution to the integral regeneration effect 
\cite{BW,LH,Rosen,BK,recent}. For small  $\tan^2 \theta$ 
the rate decreases with $\cos \theta_Z$  
and in the core one has a dip in the distribution of events.

3) The LOW solution predicts the existence of three peaks in the 
zenith angle distribution of events. There are two 
oscillation peaks in the mantle 
range: a wide peak with the maximum at $\cos \theta_Z = 0.35$ 
and a narrow peak at $\cos \theta_Z \sim 0.8$. The third peak 
(in the core region) is due to    
the parametric enhancement of oscillations. The  maximum of this peak 
is at $\cos \theta_Z \sim 0.92$.

These features are rather generic. 
The qualitative behaviour of the distribution is the same for 
all points within a given solution, although quantitatively the sizes of 
the different structures appearing in the distributions change with  
the oscillation parameters. 
Therefore, in principle a detailed study of the zenith angle distributions 
will not only give the identification of the solution of the problem
but also 
the determination of the neutrino oscillation parameters. 

Due to the relatively small present statistics (effects themselves are
small) 
it is unavoidable to bin the distribution of events. The binning should be 
chosen in such a way 
to avoid as much as possible the averaging or washing out  of the
structures. 

We propose the following binning.  
We will enumerate the night bins as 
N1 ... N5 and denote by $[\rm N]_i$,  ($i = 1...5$) the average rate 
in a given bin Ni normalized to the no oscillation prediction. 
We will call it the reduced rate.  
We will denote by [N] and [D] the averaged reduced rates 
during the night and during the day correspondingly. 

Since in the mantle range both the SMA and LMA solutions predict a rather
flat  distribution with weak dependence on $\cos \theta_Z$, 
we suggest the binning which emphasizes 
the features of the distribution for the LOW solution.  
\begin{itemize}
\item
Bin N1: $\cos \theta_Z = 0$--$0.173$ ($90^\circ>\theta_Z>80^\circ$). 
In this bin one expects very small 
regeneration effect both for the SMA and LOW solutions. Significant effect
should be observed in the case of  the LMA solution. 
\item
Bin N2: $\cos \theta_Z =  0.173$--$0.5$ 
($80^\circ>\theta_Z > 60^\circ$). This bin is selected in
such a way that it covers the main part of the wide peak of the LOW 
distribution. 
\item
Bin N3: $\cos \theta_Z = 0.5$--$0.707$
($60^\circ>\theta_Z>45^\circ$). This bin covers the dip 
of the LOW distribution: [N]$_3({\rm LOW}) \sim$ [D]. 
\item
Bin N4: $\cos \theta_Z = 0.707$--$0.83$ ($45^\circ>\theta_Z>35^\circ$)  
is the last mantle bin. This bin corresponds to the second peak of 
the LOW solution. 
%
\item
Bin N5: $\cos \theta_Z > 0.83$ ($\theta_Z <  35^\circ$), the core bin.  
This bin is restricted by  $\cos \theta_Z =  0.83$--$0.975$ for SK 
and $\cos \theta_Z =  0.83$--$0.92$ for SNO. For LMA the rate is 
comparable with the rates in the previous bins.  
For the LOW solution one expects to see a rate slightly smaller than the 
one in the fourth bin. Here the regeneration effect due to the parametric 
enhancement of oscillations takes place. For SMA the effect depends 
strongly on the oscillation
parameters: for large $\tan^2 \theta$ one should see the highest 
rate. With decrease of the mixing angle the rate decreases
and becomes smaller that the day rate. 
\end{itemize}

In Figs.~\ref{bin-lma},~\ref{bin-sma}, and~\ref{bin-low} we show the binned
event rates for [ES] events  at SK and [CC] events at SNO for oscillation 
parameters in the LMA, SMA and LOW solutions respectively.  
The distributions are rather similar for 
[ES] events at SK and [CC] events at SNO. 
Although the enhancement of the Earth--matter effect in the core bin 
for the SMA solution is larger in SK, since this bin covers a larger
interval 
of $\cos \theta_Z$ at the SK latitude. 

The qualitative  behaviour 
of the [CC] rates does not change with the increase of the threshold energy 
from 5 MeV to 8 MeV for SNO. The absolute value of the effect 
(reduced rate), however, 
increases for the LMA solution and it decreases for the LOW and SMA solutions. 

For the LMA solution the regeneration
effect decreases with the increase of $\Delta m^2$ in all the bins and it 
changes weakly with the mixing. 
For the LOW solution the situation is opposite: the rates  
[N]$_2$, [N]$_4$ 
and [N]$_5$ increase with $\Delta m^2$.  This behaviour originates from
the dependence of $\freg$ on $\Delta m^2$ (see Eqs.~(\ref{regen-low}) and
(\ref{peak-param})). Also for SMA  the rate in the core bin
increases with $\tan^2 \theta$. 

\begin{figure}[!t]
\centerline{\psfig{figure=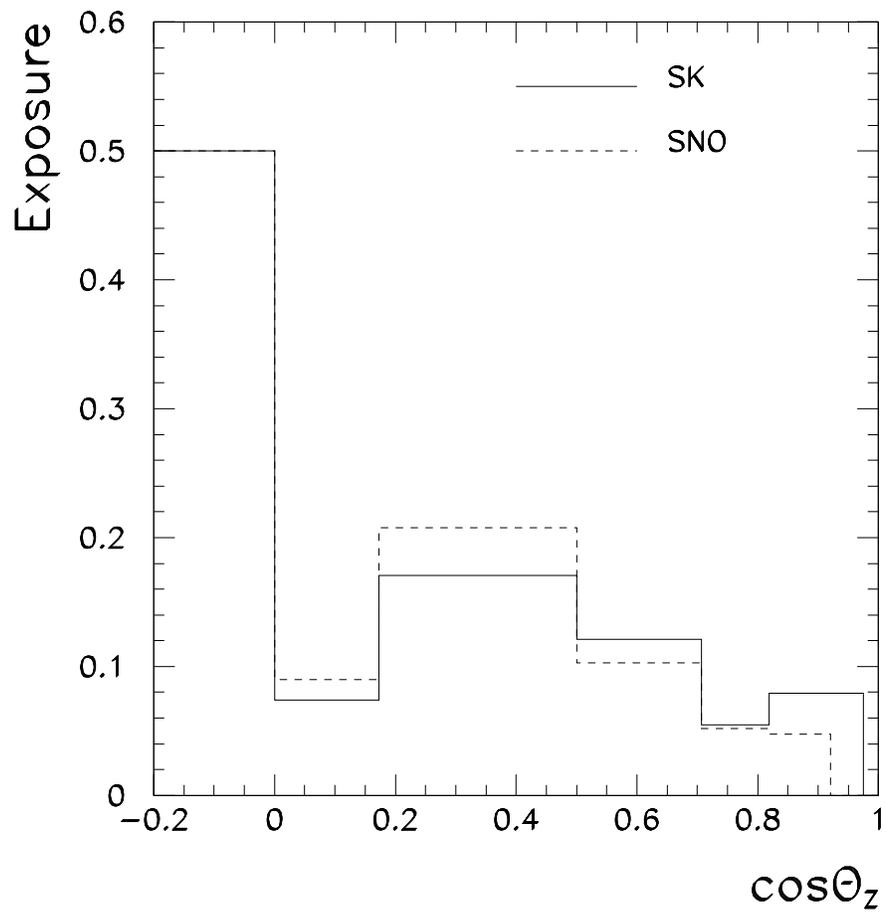,width=5.0in,angle=0}}
\tightenlines  
\caption{The exposure time for the different $\cos \theta_Z$ bins 
for SNO and SK detectors during the year.}
\label{time}
\end{figure}

The comparison of the event 
rates rates in the different bins allows to discriminate among the 
solutions of the solar neutrino problem. In particular, for LMA we have 
\begin{equation}
{\rm [N]}_1 < {\rm [N]}_2 \leq   {\rm [N]}_3
\leq {\rm [N]}_4 \leq {\rm [N]}_5.
\label{ineqlma}
\end{equation}
For LOW distribution: 
\begin{equation}
{\rm [N]}_2 \geq {\rm [N]}_4 > {\rm [N]}_1 \sim {\rm [N]}_3 > 
{\rm [D]}.
\label{ineqlow} 
\end{equation}
For SMA distribution:   
\begin{equation}
|{\rm [N]}_1 - {\rm [D]}| <  
|{\rm [N]}_2 - {\rm [D]}| \leq
|{\rm [N]}_3 - {\rm [D]}| \leq
|{\rm [N]}_4 - {\rm [D]}| \leq
|{\rm [N]}_5 - {\rm [D]}|.  
\label{ineqsma} 
\end{equation}
Finally, for the sake of completeness we show in Fig.~\ref{time} 
the time exposure for the different bins (average during the
year) for the SK and SNO detectors. The histograms are normalized in such
a way that the integral over night time gives 0.5. The time exposure
distribution determines the statistics in each bin.  
As follows from the figure the highest statistics is expected in N2 
followed by N3. Moreover, at SNO the time exposure  N2 will be about twice
longer than N3,  
while for SK the difference in statistics is smaller.  
For SNO the lowest statistics is expected in  N5, whereas for 
SK it corresponds to N4.

\subsection{Correlations}
\label{subsec:corr}

At present, the expected regeneration effects are restricted by the 
Super--Kamiokande result on the day--night asymmetry: 
\begin{equation} 
A_{\rm{DN}}^{\rm SK} = 2 \frac{N-D}{N+D}\sim 0.034\pm 0.026~.    
\end{equation}
The asymmetry gives the integrated (over $\cos \theta_Z$) Earth--matter
effect. With present statistics the integral effect is at $1.3 \sigma$ 
level which means that in spite of large number of events 
it will be difficult  to measure the zenith angle distribution with  a
high precision. Higher asymmetry (due to absence of damping) is expected for 
SNO \cite{bkssno}: 
\begin{equation}
A_{{\rm ND}}^{\rm SNO} <  0.10 - 0.15. 
\end{equation}

\begin{figure}[!t]
\centerline{\psfig{figure=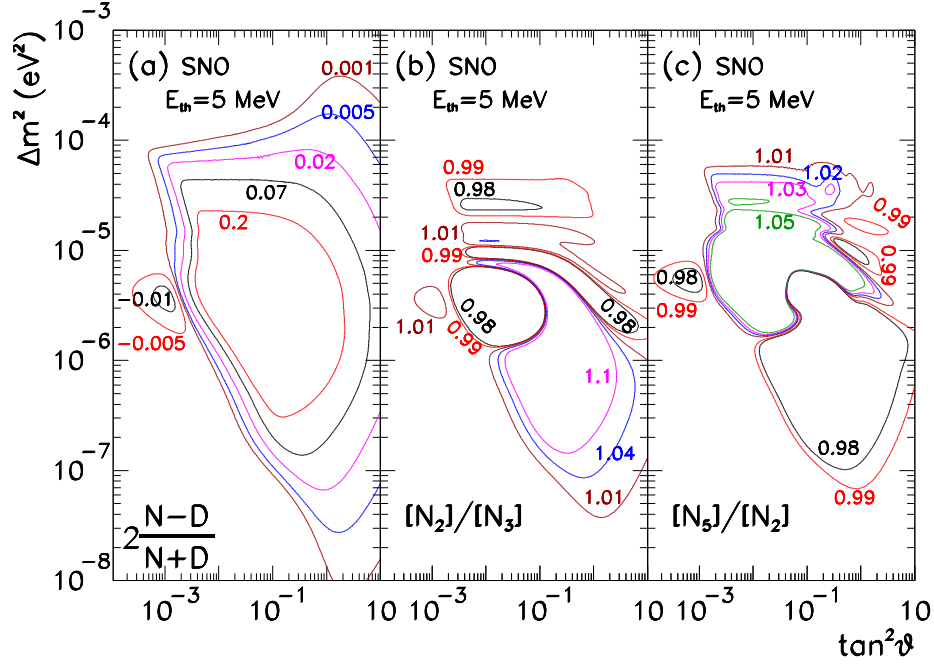,width=5.0in,angle=0}}
\tightenlines  
\caption{The contours of equal Day-Night asymmetry, $A_{\rm{DN}}$ (a), 
the ratios of rates $N_5/N_2$ (b), and $N_2/N_3$ (c) at  SNO 
( threshold 5 MeV ) in the $\Delta m^2 - \tan^2 \theta$ plane.}
\label{contours}
\end{figure}

To enhance the identification power of the analysis 
one can study the correlations between the signals in different zenith angle
bins 
and the integral effect which can be represented by the 
day--night asymmetry. As an illustration 
of such  correlations  we first show in  Fig.~\ref{contours}
contours of constant asymmetry $A_{\rm{DN}}$ and of ratios 
of the reduced rates [N]$_2$/[N]$_3$ and
[N]$_5$/[N]$_2$ in the 
$\Delta m^2 -  \tan^2 \theta$ plane for SNO with threshold 5 MeV. 

For the LMA solution the $\cos \theta_Z$ - distribution is rather flat 
which means that for all bins 
\begin{equation}
\frac{\rm [N]_i}{\rm [N]} - 1 \propto A_{\rm{DN}}. 
\label{ni-n-lma}
\end{equation}

For the LOW solution the heights of all three peaks increase
with integral effect: 
\begin{equation}
\frac{\rm [N]_2}{\rm [N]} - 1 \propto A_{\rm{DN}}
\label{ni-n-low}
\end{equation}
and similar proportionality exists for ${\rm [N]_4}$ and 
${\rm [N]_5}$. 
In the third bin, however, the regeneration effect is practically zero and 
the rate is at the level of the day rate.

For the SMA solution, the rate monotonously increases 
(or decreases depending on $\tan\theta$) with 
$\cos \theta_Z$,  so that for all the bins the proportionality 
(\ref{ni-n-lma}) holds approximately.

\subsubsection{Ratio ${\rm [N]_5/[N]_2}$  versus $A_{\rm{DN}}$.  }
Comparing  the signals in N2 and N5 bins we get 
\begin{equation}
\frac{\rm [N]_5}{\rm [N]_2} \sim 1 ~~~ ({\rm LMA}), ~~~~~~~~~~  
\frac{\rm [N]_5}{\rm [N]_2} 
> 1 ~~~~ ({\rm LOW})~.   
\label{ratio5-2}
\end{equation}
In both these cases the day--night asymmetry is positive. On the other hand 
for SMA we have ${\rm [N]_5/[N]_2} > 1$ for the part of the 
allowed region with
small mixing, for   $A_{\rm{DN}} < 0$,
and  ${\rm [N]_5/[N]_2} <  1$ for the part of the region with large
mixing, where  $A_{\rm{DN}} > 0$  (see Fig.~\ref{contours}).

\begin{figure}[!t]
\centerline{\psfig{figure=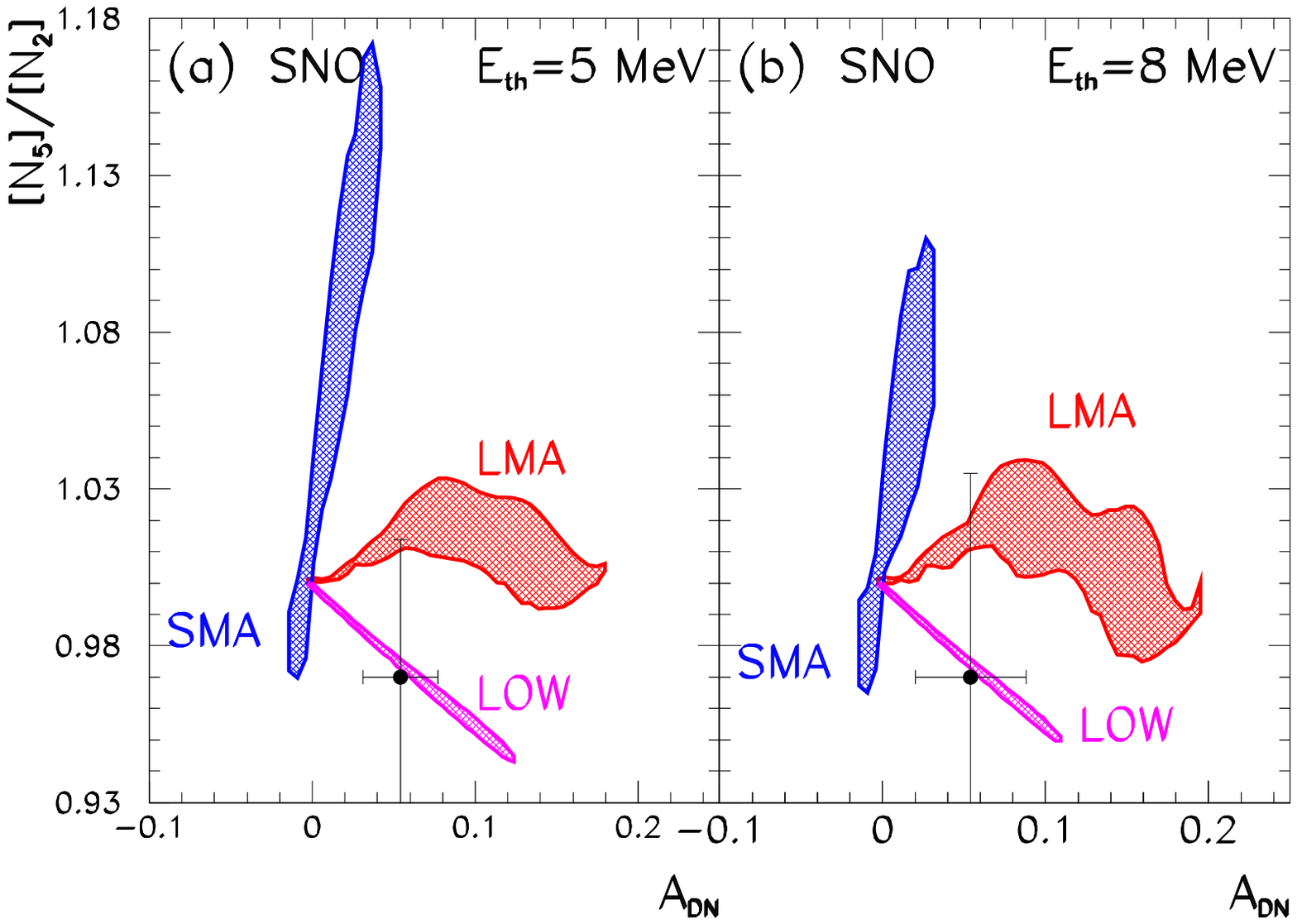,width=5.0in,angle=0}}
\tightenlines  
\caption{The allowed values for the day--night asymmetry, 
$A_{\rm{DN}}$ and the ratio of rates in the N5 and N2 bins, $N_5/N_2$,  at 
SNO above 5 MeV (a) and 8 MeV (b) in the different allowed regions 
(at 99\% CL) of the solar neutrino problem. 
We also 
show the expected statistical sensitivity after 3 years of operation 
(see text for details). }
\label{sno-5/2}
\end{figure}

\begin{figure}[!t]
\centerline{\psfig{figure=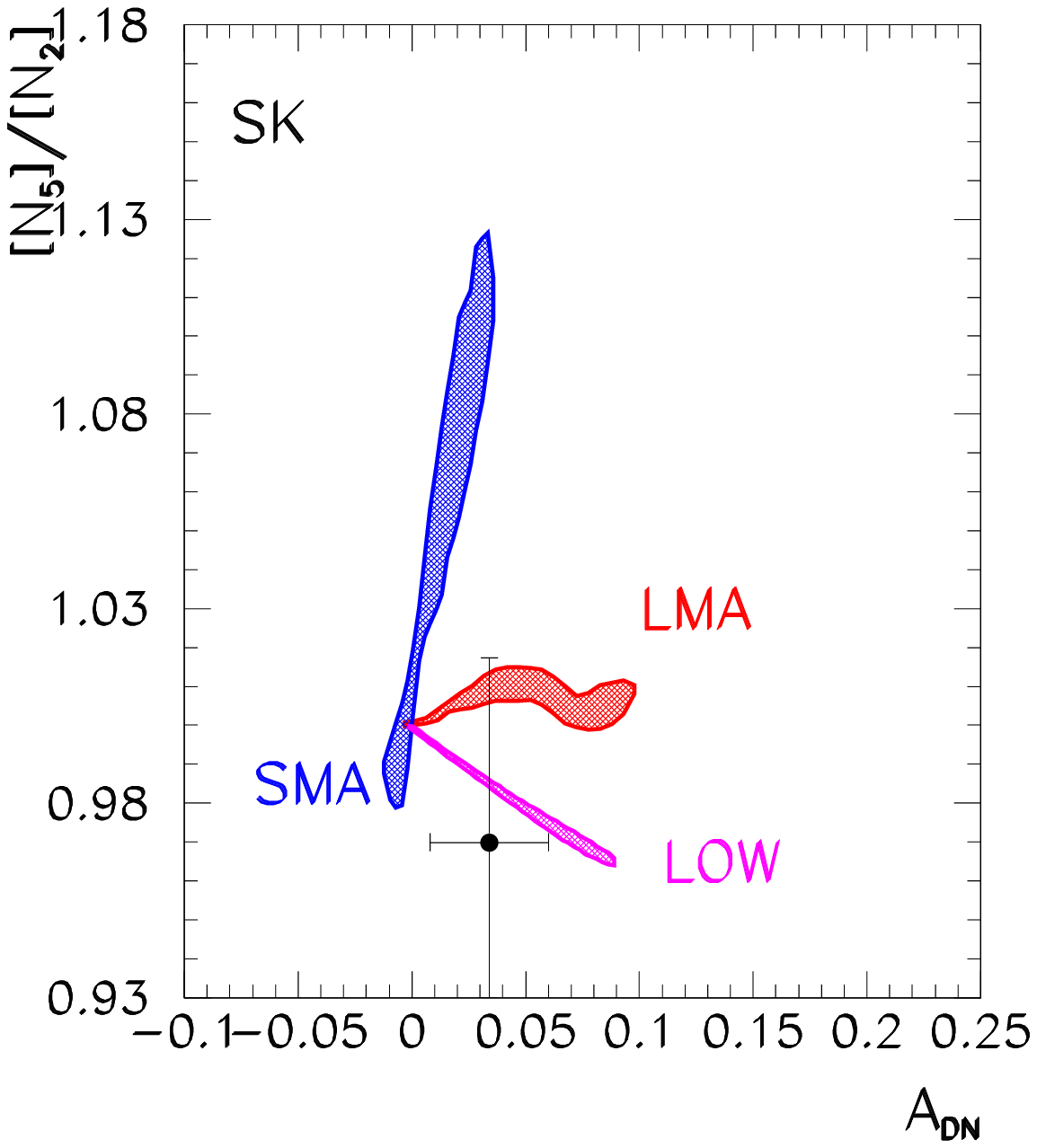,width=5.0in,angle=0}}
\tightenlines  
\caption{The allowed values for the day--night asymmetry, 
$A_{\rm{DN}}$ and the ratio of rates in the N5 and N2 bins, $N_5/N_2$,  
at SK in the different
allowed regions (at 99\% CL) of the solar neutrino problem.
The present experimental values and 1 $\sigma$ errors are also displayed
(see text for details).}
\label{sk-5/2}
\end{figure}

In Figs.~\ref{sno-5/2}, and~\ref{sk-5/2} we show the results of mapping 
the 99\% CL solution regions onto the plane 
of the ${\rm [N]_5/[N]_2} - A_{\rm{DN}}$ 
observables for the SNO and SK experiments. 
From the figures we see that there is a strong correlation 
between the ${\rm [N]_5/[N]_2}$ ratio and the asymmetry for all three
solutions. 

For SNO observables above 5 MeV (Fig.~\ref{sno-5/2}) the correlation can be
parametrized as: 
\begin{equation}
\frac{\rm [N]_5}{\rm [N]_2}  \approx \left\{
\begin{array}{ll}
1             &  {\rm ~~~~~ LMA} \\
1 - 0.45 A_{\rm{DN}}   &  {\rm~~~~~ LOW}\,  \\ 
1 + 3.6 A_{\rm{DN}}    &  {\rm ~~~~~SMA } 
\end{array}
\right. . 
\label{corr25}
\end{equation}
A negative asymmetry would testify for the SMA solution  but the maximal
asymmetry which is expected for the SMA on the basis of the present data
cannot be larger than 0.05. Thus, for 
$A_{\rm{DN}}({\rm SNO})  > 0.05$ (above 5 MeV)  one will 
choose between LMA and LOW. A ratio  
${\rm [N]_5/[N]_2} > 1$ will testify for LMA,
whereas  ${\rm [N]_5/[N]_2} <  1$ will be an evidence of the LOW solution. 
For $A_{\rm{DN}} = 0$, the equality ${\rm [N]_5/[N]_2} = 1$ is fulfilled 
for all the solutions. 
In the range  $A_{\rm{DN}}({\rm SNO}) <  0.05$ all three solutions are
possible. 
With increase of  the asymmetry the difference 
of ratios for different solutions increases.  

For the SNO measurements  with $E_{th} = 8$ MeV, 
the correlations are described approximately by 
(\ref{corr25}), but the LMA region has a more irregular shape since at 
$E_{th} = 8$ MeV the averaging is weaker. 
The maximal allowed deviations of ${\rm [N]_5/[N]_2}$ 
from  1 are $\sim 11\%$ ({\rm SMA}), $\sim 4\%$ ({\rm LMA}) 
and $\sim - 5\%$ ({\rm LOW}). 

Comparing with the results at lower energy threshold we find that  
the allowed SMA region in  ${\rm [N]_5/[N]_2} - A_{\rm{DN}}$ plane  
shrinks and $A_{\rm{DN}}({\rm SNO})$ can reach only 0.04,  
while ${\rm [N]_5/[N]_2}$ can still be as large as 1.11. 
This happens because the region of parametric enhancement is less  
covered with the higher threshold. 
For the LOW solution the allowed  region increases (since the regeneration 
effect decreases with energy). $A_{\rm{DN}}({\rm SNO})$ can reach 11\% and 
$({N_5}/{N_2} - 1) \sim 0.05$. 

For the ES - events at SK (Fig.~\ref{sk-5/2}) one gets a similar picture, 
although  the regeneration effects are damped by the
$\nu_{\mu}$ and $\nu_{\tau}$ 
contribution. The expected maximal asymmetry 
is $A_{\rm DN} (SK) \sim 3 \%$ for SMA and $A_{\rm{DN}}({\rm SK}) 
\sim 10 \%$
for LMA and LOW.  If  $A_{\rm{DN}}({\rm SK}) \sim 5 \%$,  
the ${\rm [N]_5/[N]_2}$ 
ratio for  LMA is larger than  
the ratio for LOW  by $(0.03 - 0.05)$.   
 
To have an idea about present sensitivity in searches of correlations, 
we also show in Fig.~\ref{sk-5/2} 
the SK results and with  1~$\sigma$ 
errors for the day--night
asymmetry and the ${\rm [N]_5/[N]_2}$ ratio. One must bare in mind,
however,  that the plotted result on 
${\rm [N]_5/[N]_2}$ does not correspond to the
binning we are proposing. 
In Fig.~\ref{sno-5/2} we show the corresponding attainable SNO sensitivity 
for three years of operation (corresponding to 13200 CC events for  
5 MeV threshold and 6000 CC events for 8 MeV threshold). For definiteness we 
have plotted the 
point at a value $A_{\rm{DN}}({\rm SNO}) =1.6 A_{\rm{DN}}({\rm SK})$
which is the expected relation for the best fit point in LMA as
discussed in Ref.~\cite{bkssno}.

\subsubsection{Ratio ${\rm [N]_2/[N]_3}$  versus $A_{\rm{DN}}$.  }

According to Eqs.~(\ref{ineqlma})--(\ref{ineqsma}) 
(see also Fig.~\ref{contours}), we have 
\begin{equation}
{\rm [N]_2} \approx {\rm [N]_3 } ~~~ ({\rm  LMA}), ~~~
{\rm [N]_2} > {\rm [N]_3 }  ~~~ ({\rm LOW}), ~~~
|{\rm [N]_2} - {\rm [D]}| <  |{\rm [N]_3 } - {\rm [D]}|~~~ ({\rm SMA})~.  
\label{rel2-3}
\end{equation}
So that the rates in the high statistics bins N2 and N3 can provide 
an important criteria to distinguish the solutions. 

\begin{figure}[!t]
\centerline{\psfig{figure=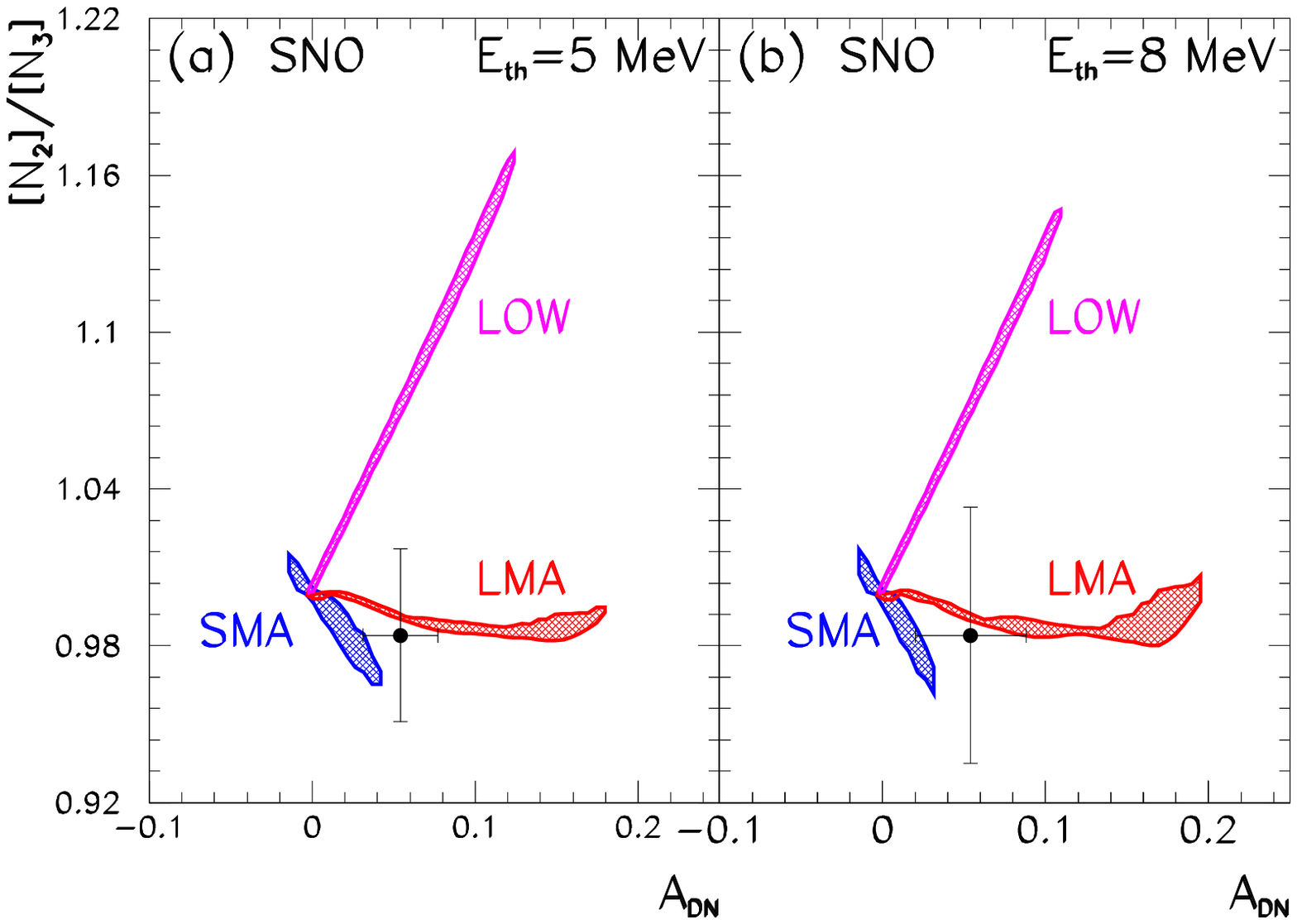,width=5.0in,angle=0}}
\tightenlines  
\caption{The allowed values for the day--night asymmetry, 
$A_{\rm{DN}}$ and the ratio of rates in the N2 and N3 bins, $N_2/N_3$,  at 
SNO above 5 MeV (a) and 8 MeV (b) in the different allowed regions 
(at 99\% CL) 
of the solar neutrino problem.
We also 
show the expected statistical sensitivity after 3 years of operation 
(see text for details).}
\label{sno-2/3}
\end{figure}

\begin{figure}[!t]
\centerline{\psfig{figure=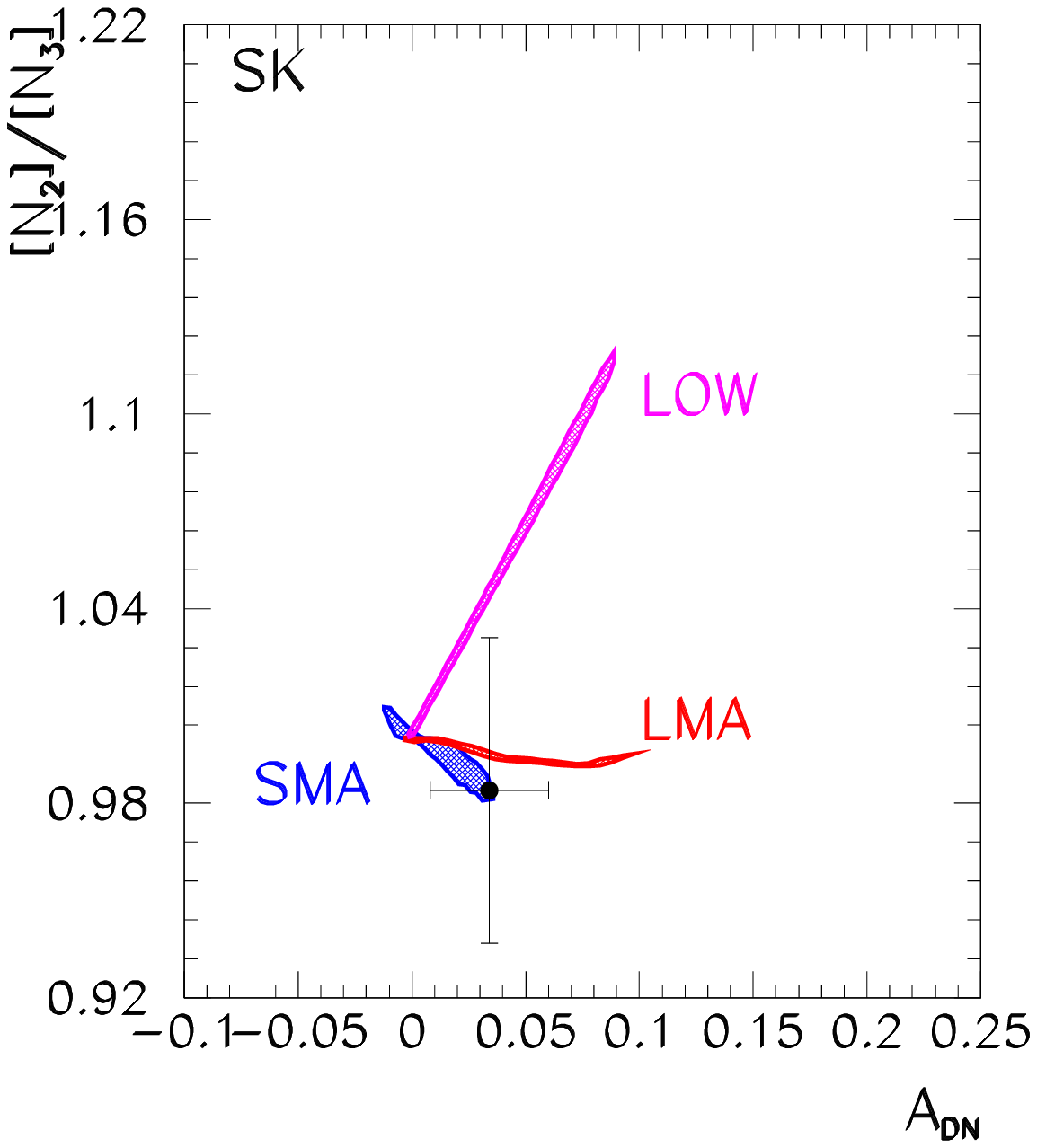,width=5.0in,angle=0}}
\tightenlines  
\caption{The allowed values for the day--night asymmetry, 
$A_{\rm{DN}}$ and the ratio of rates in the N3 and N2 bins, $N_3/N_2$,  
at SK in the different
allowed regions (at 99\% CL) of the solar neutrino problem.
The present experimental values and 1 $\sigma$ errors are also displayed
(see text for details).}
\label{sk-2/3}
\end{figure}

In Fig.~\ref{sno-2/3} we show the mapping the 99\% CL allowed
regions of oscillation parameters onto the ${\rm [N]_2/[N]_3} -
A_{\rm{DN}}$ plane 
of the SNO observables for two different energy thresholds. 
From the figure one can see that there is a clear correlation between  
the ratio and the asymmetry for the SMA and LOW solutions: 
\begin{equation}
\frac{\rm [N]_2}{\rm [N]_3}  \approx \left\{
\begin{array}{ll}
1        &  {\rm ~~~~~ LMA} \\
1 + 1.3\, A_{\rm{DN}}   &  {\rm ~~~~~ LOW} \\
1 - 1.1\,  A_{\rm{DN}}  &  {\rm ~~~~~ SMA}
\end{array}
\right. . 
\label{corr23}  
\end{equation} 
Also, we see that the regions only weakly depend on the energy threshold. 
In the case of the LOW solution 
the  ${\rm [N]_2/[N]_3}$ - ratio can reach 1.17 for $E_{th} = 5$
MeV  and  1.15 for $E_{th} = 8$ MeV.  
For SMA the maximal allowed deviation of the ${\rm [N]_2/[N]_3}$ ratio
from 1 is about 4\%. 
We get qualitatively similar results for SK (Fig.~\ref{sk-2/3}), 
where the regeneration effect is damped by $\nu_{\mu}$ and $\nu_{\tau}$ 
contributions, so that the allowed regions for all three solutions are 
smaller than for SNO. We also plot in 
Figs.~\ref{sk-2/3} and~\ref{sno-2/3} the present results from SK
and the estimated accuracy at SNO as discussed at the end of 
the  previous section.

From these results it is clear that, with the estimated  statistics 
at SK and SNO, it will be difficult to get strong discrimination among the 
solutions. At $3 \sigma$ level all the solutions 
seem to be allowed. However, future higher statistics experiments  
like UNO \cite{UNO} can give a more significant  result. 

One may also attempt  to elaborate a test criteria for some of the  solutions 
without a large lost of statistics. For instance, for the LOW solution 
one expects a significant deviation from one for the ratio:   
\begin{equation}
\frac{{\rm [N]_2} + {\rm [N]_4} + {\rm [N]_5}}{{\rm [N]_1} +  
{\rm [N]_3}} > 1.  
\end{equation}
While for LMA this ratio should be close to one.


\section{Conclusions}
\label{sec:conclusion}

In this work we have studied in detail the zenith
angle dependence of the regeneration factor and the distributions of the
solar neutrino events expected in 
the Super--Kamiokande and SNO detectors. 

We have  identified  the generic features of the distributions 
for each solution. 

1. The LMA $\cos \theta_Z$ - distribution is characterized by a fast
oscillatory behavior with rather slowly changing averaged 
regeneration rate. The amplitude of oscillations is enhanced in the  
core region.  
The behaviour of the event rates at SNO and SK has an interesting 
``synchronization effect" so that  strong oscillations of the rates 
with the change of 
$\cos \theta_Z$ are only expected at small and at large values of 
$\cos \theta_Z$. In particular 2 - 3 oscillations 
with significant amplitude are expected  at small $\cos \theta_Z$. 

To detect such a behavior one needs to use small size
binning 
$\Delta (\cos \theta_Z)  \sim   0.05$ which (in view of relatively low
statistics) may only be possible with new megaton water Cherenkov
detectors. 
If the LMA solution is identified, a study of the oscillatory
behavior can be used to measure the oscillation parameters 
(in particular $\Delta  m^2$) and the Earth density profile.

2. The SMA solution predicts a smooth zenith angle dependence in the
mantle region with small regeneration effect while  the  
parametric effects can show up in the core region. 
For the larger mixing part of the allowed region 
the regeneration effect is positive with a parametric peak in the core 
region; for the smaller mixings the effect is negative and in the core 
region one should see the parametric dip.

3. In the case of the LOW solution there are two oscillation peaks in the 
mantle region with maxima at $\cos \theta_Z = 0.35$ 
(oscillation phase $\phi = \pi$) and at 
$\cos \theta_Z = 0.78$ (the phase $3 \pi$). 
In the core region the third peak appears due to 
parametric enhancement of oscillations. In the maximum 
the conditions for the parametric resonance are approximately satisfied. 
This realization of the parametric resonance corresponds to the case of 
neutrino energy above the resonance energy in the mantle 
(which differs from the realization for the SMA) 
and to the phase relations $\phi_{mantle} \approx \pi$ and 
$\phi_{core} \approx 3 \pi$. 
We find that the position of the peaks is almos independent  
of the neutrino energy, whereas the height of the peaks as well as the 
integral effect are inversely proportional to the energy.
 
The shape of the distributions and their dependence on the oscillation
parameters as well as  energy thresholds are well described by
oscillations 
in a medium which consists of one or three layers with slowly 
changing 
density. Such an oscillation picture gives the correct functional dependence
of the rates on the neutrino parameters.

We also show that precise measurements  of the zenith angle distribution will
allow to study a number of various matter effects such as
oscillation in matter with constant and slowly changing density, 
adiabatic conversion, parametric enhancement of oscillations, etc\dots.

We have suggested a new binning in the zenith angle distribution 
which is chosen to emphasize  the distinctive features of the 
distributions for the different solutions. The identification power of 
the analysis can be 
further enhanced by studying correlations between the event rates in the 
different bins and the integral 
effect which can be represented by the day--night asymmetry. 
In particular we have shown that the study of the correlation  
between the ratio ${\rm [N]_2/[N]_3}$ and the day--night asymmetry 
will help   to identify  the LOW solution while
the correlation between ${\rm [N]_5/[N]_2}$ and $A_{\rm D/N}$ is more
suitable for identification of  the SMA solution. 
The distributions expected at SNO and at SK have similar 
properties. But at SNO the structures in the distributions and the 
asymmetry are enhanced due to the absence of the damping effect.

\acknowledgments
This work was supported by the spanish DGICYT under grants PB98-0693 
and PB97-1261, by the Generalitat Valenciana under grant
GV99-3-1-01 and by the TMR network grant ERBFMRXCT960090 of the 
European Union.

\section*{Appendix. Graphic representation of evolution of the neutrino
state}
 
We describe here a graphic representation of the evolution of the neutrino 
state
which allows us to understand the properties of the zenith angle distributions 
for the different solutions. The graphic representation \cite{KS} is based
on the analogy of the neutrino evolution with the behaviour of
spin of the electron in the magnetic field. The 
neutrino state is  described by a vector of length 1/2 with components 
\begin{equation}
\vec{\nu} = \left( {\rm Re} \psi_{e}^{\dagger} \psi_{\mu}, ~~
{\rm Im} \psi_{e}^{\dagger} \psi_{\mu},~~
\psi_{e}^{\dagger} \psi_{e} - 1/2 \right) ~,
\label{neutvec}
\end{equation}
where $\psi_{i}$, ($i = e,  \mu$) are the neutrino wave functions 
of the electron and muon neutrinos. 
(The components  of this vector are  elements  of the
density matrix.)

Introducing  the vector 
\begin{equation}
\vec{B} \equiv \frac{2 \pi}{l_m} (\sin 2 \theta_m,~~ 0 ,~~ \cos 2
\theta_m)~
\label{axisb}
\end{equation}
($\theta_m$ is the mixing angle and $l_m$ is the oscillation length in the 
medium) which would correspond to the
magnetic field,  one can get from the 
evolution equation for $\psi_{i}$  the equation for $\vec{\nu}$: 
\begin{equation}
\frac{d \vec{\nu}}{d t} = \left(\vec{\nu} \times \vec{B} \right)~.
\end{equation}

According to (\ref{neutvec}) the projection  of $\vec{\nu}$ on the axis
$z$, $\nu_{z}$, gives the probability
to find $\nu_{e}$ in the state $\vec{\nu}$:
\begin{equation}
P \equiv \psi_{e}^{\dagger} \psi_{e} =
\nu_z + \frac{1}{2} =
\cos^2 \frac{\theta_e}{2} ~.
\label{thetaz}
\end{equation}   
Here $\nu_{z} \equiv 0.5 \cos \theta_e$,
and  $\theta_e$ is the angle between $\vec{\nu}$ and the axis $z$.
The $z$ axis can be called the flavor axis as it coincides with 
vectors $\vec{\nu}_e = - \vec{\nu}_{\mu}$. 

\begin{figure}[!t]
\centerline{\psfig{figure=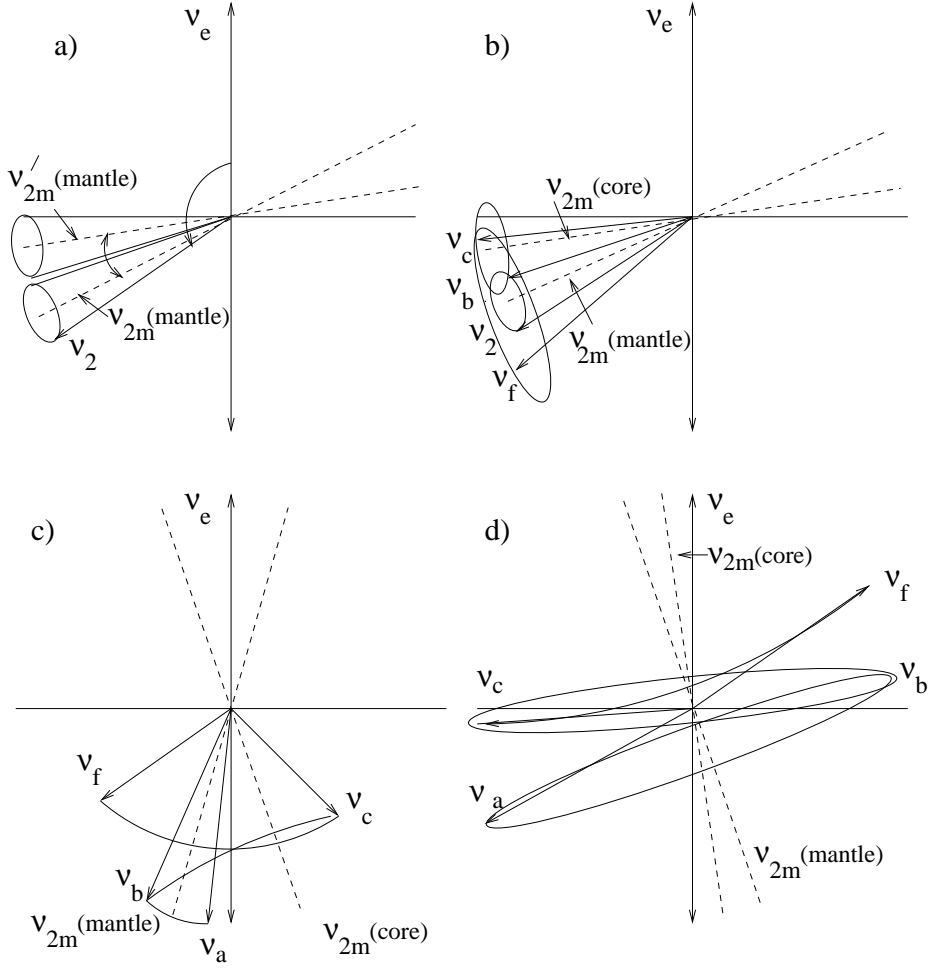,width=5.0in,angle=0}}
\tightenlines  
\vskip 1.cm
\caption{Graphic representation of the evolution of the neutrino state
inside  the Earth  for different solutions of the solar neutrino problem. 
(a) LMA solution: propagation along trajectory 
which crosses the mantle only; 
(b) LMA solution: evolution for core crossing trajectory; 
(c) SMA solution: evolution for core crossing trajectory 
and $\cos \theta_Z$ which corresponds to the maximum of the parametric
peak; (d) the same as in (c) for LOW solution.}
\label{cones}
\end{figure}

In a medium with constant density, $\theta_m = const$,
the evolution corresponds to the precession of the vector $\vec{\nu}$ -
around $\vec{B}$:  $\vec{\nu}$  moves
according to the increase of the oscillation phase, $\phi$,
on the surface of the cone with axis $\vec{B}$.
The vector $\vec{B}$ coincides with the vector of the eigenstates in
matter $\vec{\nu}_{1m} = -\vec{\nu}_{2m}$. 
We will denote by $\vec{\nu}_{2m}(mantle)$ 
and $\vec{\nu}_{2m}(core)$  
the axis in the mantle and in the core. 
Notice that when the matter
density is below the resonant one ($\eta>1$) the corresponding axis  
lies in the first-third quadrants with decreasing projection
on the $z$ axis as the density increases (see Figs.~\ref{cones}a 
and~\ref{cones}b). Once the resonant density
is crossed, the axis moves to the second-fourth quadrants 
(see Figs.~\ref{cones}c and~\ref{cones}d).


In Figs.~\ref{cones}a and  Figs.~\ref{cones}b 
we show the evolution of the neutrino state 
in the case of the LMA solution for trajectories crossing the mantle 
only (a) or the mantle and the core (b). Inside the Sun the neutrino vector 
(produced as $\vec{\nu} = \nu_e \approx  \vec{\nu}_{2m}$ follows the vector 
$\vec{\nu}_{2m}$ due to adiabaticity and appears at the surface of the Sun
and then at the surface of the Earth as $\vec{\nu} = \vec{\nu}_{2}$. 
Due to the adiabatic density change  in the mantle the axis 
$\vec{\nu}_{2m}$ moves accordingly,
but at the detector it (as well as the cone of precession) will return to
the same position as in the moment of arrival  of the neutrino at the
Earth.  The cones with axis $\vec{\nu}_{2m}$ and  $\vec{\nu}_{2m}'$ 
describe  oscillations at the
surface of the Earth and in the central part of the trajectory
correspondingly.  

At the surface of the 
Earth the density jumps suddenly and therefore the axis vector 
$\vec{\nu}_{2m}$ suddenly  changes its position to from $\vec{\nu}_{2}$ 
to  $\vec{\nu}_{2m}(mantle)$. In the mantle the neutrino vector 
will precess around $\vec{\nu}_{2m}(mantle)$. Clearly, since
$\vec{\nu}_{2m}(mantle)$ is closer than $\vec{\nu}_{2}$ to $\nu_e$, 
the rotation will lead to an increase of the $z$ projection 
and therefore to increase the $P_{ee}$ probability. 
This illustrates the fact that the regeneration is always positive 
in the mantle zone (see also \cite{wolf}). 
The quasi-adiabatic movements
of $\vec{\nu}_{2m}(mantle)$ due to density variations in the mantle
do not change this conclusion. 

Let us now consider the evolution for the core crossing trajectory. 
The neutrino vector starts at $\vec{\nu}_a = \nu_2$ and
it will first precess around $\vec{\nu}_{2m}(mantle)$. The phase 
of oscillations  is rather large  $\sim (10 - 20) \pi$ and it  depends 
sensitively on  
the neutrino energy and the zenith angle. So, depending on the energy the 
neutrino state can arrive at the core 
in any position  on the cone.  In Fig.~\ref{cones}b, as an example,  we have
selected the state $\vec{\nu}_b$. In the core the  vector 
$\vec{\nu}$ will precess around $\vec{\nu}_{2m}(core)$ starting from the
position $\vec{\nu}_b$. The phase is large and the state will enter the
mantle again 
in some position $\vec{\nu}_c$ which depends on $E$ and $\theta_Z$. 
In the second mantle layer the state will precess around
$\vec{\nu}_{2m}(mantle)$  starting from $\vec{\nu}_c$ state. 
Let us denote by $\vec{\nu}_f$ the final neutrino state at the detector. 
As it is clear from the figure, in most of configurations determined  
by the positions of $\vec{\nu}_c$ and $\vec{\nu}_b$ (and therefore $E$ and
$\theta_z$) the final state will have the projection $\vec{\nu}_z$ 
larger that the projection of $\vec{\nu}_a = \nu_2$ (which corresponds to 
day signal). In other words: 
$\cos 2\theta_e (\vec{\nu}_f) <  \cos 2\theta$. 
In some cases, however, one can get $\nu_z (final) < \nu_{2z}$ which
corresponds to negative regeneration.  

It is easy to see that maximal regeneration effect would correspond 
to zero regeneration effect in the mantle: $\phi_{mantle} = 2\pi k$ and 
to maximal effect in the core: $\phi_{core} = \pi + 2 \pi k'$ 
($k, k'$ are integer numbers). 

In Fig. \ref{cones}c we show the  evolution of the neutrino state in the 
case of SMA solution. Now  the neutrino arrives at the Earth as 
an incoherent superposition of $\nu_1$ and $\nu_2$. This split of the state
originates from the adiabaticity violation inside the Sun.   
The total survival probability 
$P_{ee}$ is then determined  by independent oscillations of 
$\nu_1$ and $\nu_2$ inside the Earth. The result can be represented 
in terms on $\nu_2 \rightarrow \nu_e$ probability and the adiabaticity
factor (\ref{Pday}). The later can have both negative and positive signs 
depending on the oscillation parameters, and it is this factor 
which determines the sign of the regeneration effect. 

In Fig.~\ref{cones}c  we show the evolution of the $\nu_2$ state inside 
the Earth. In the mantle the vector $\vec{\nu}$ precesses around
$\vec{\nu}_{2m}(mantle)$ starting from $\vec{\nu}_a = \nu_2$. 
In the core region, the parametric enhancement of oscillations 
occurs for certain values of $\cos \theta_Z$. 
We show the graphic representation of evolution for this case. 
Neutrino arrives at the core in
the state  $\vec{\nu}_b$, then  $\vec{\nu}$  precesses around
$\vec{\nu}_{2m}(core)$ starting from $\vec{\nu}_b$
and reaches the border of core and mantle in the state   
$\vec{\nu}_c$ (which corresponds to the phase  $\phi_{core}\sim 0.75\pi$).  
In the second mantle layer $\vec{\nu}$ rotates around
$\vec{\nu}_{2m}(mantle)$ starting from $\vec{\nu}_c$
and arrive at the detector in the state 
$\vec{\nu}_f$. Clearly, the regeneration turns out to be enhanced.

In the case of the LOW solution (Fig.~\ref{cones}d), 
$\vec{\nu}$ arrives  at the
Earth as $\vec{\nu}_a \approx \nu_2$ similarly to the LMA case but 
in contrast to LMA, now the  mixing is matter suppressed so that the 
axis of the eigenstates is close to the flavor axis ($z$). In the mantle
$\vec{\nu}$ precesses around  $\vec{\nu}_{2m}(mantle)$ starting from
position $\vec{\nu}_a$. The peaks in the zenith angle distribution
in the mantle region correspond to the final position of the neutrino 
$\vec{\nu}_b$. In the core region the parametric enhancement of
oscillations occurs. The maximum of the parametric peak corresponds to the
following picture. 
Vector $\vec{\nu}$  precesses in the mantle and
reaches the state $\vec{\nu}_b$ at the border of  the core 
($\phi_{mantle} \approx  \pi$).  In the core $\vec{\nu}$ rotates around 
$\vec{\nu}_{2m}(core)$ starting from $\vec{\nu}_b$. 
The neutrino vector make 1.5 turns ($\phi_{core} = 3 \pi$) and leave the
core in the position $\vec{\nu}_c$. In the second mantle layer 
$\vec{\nu}$ rotates again around $\vec{\nu}_{2m}(mantle)$ 
with initial position $\vec{\nu}_c$. At the detector one will  
detect the state which corresponds to $\vec{\nu}_f$.


\newpage
\vskip 2cm
%
\vskip 2cm
%
\vskip 2cm
\vskip 2cm
\vskip 2cm
\vskip 2cm
\vskip 2cm
\vskip 2cm
\vskip 2cm
\vskip 2cm
\vskip 2cm
\vskip 2cm
\vskip 2cm
\vskip 2cm
\vskip 2cm
\vskip 2cm

\end{document}